\documentclass[a4paper,fleqn,usenatbib]{mnras}

\usepackage{newtxtext,newtxmath}

\usepackage[T1]{fontenc}
\usepackage{ae,aecompl}

\usepackage{graphicx}	
\usepackage{amsmath}	
\usepackage{amssymb}	

\newcommand{\thetaE}{$\theta_{\mathrm{E}}$}
\newcommand{\msun}{$M_\odot$}

\newcommand{\eg}{{e.g.},\,}
\newcommand{\ie}{{i.e.},\,}

\title[Microlensing black hole masses from Gaia and OGLE]{On the accuracy of mass measurement for microlensing black holes as seen by Gaia and OGLE}

\author[K. A. Rybicki et al.]{Krzysztof A. Rybicki,$^{1}$\thanks{E-mail: krybicki@astrouw.edu.pl}
{\L}ukasz Wyrzykowski,$^{1}$
Jakub Klencki,$^{1,2}$
Jos de Bruijne,$^{3}$
\newauthor
Krzysztof Belczy{\'n}ski,$^{4}$
Martyna Chru{\'s}li{\'n}ska$^{1,2}$
\\
\\
$^{1}$Warsaw University Astronomical Observatory, Aleje Ujazdowskie 4, 00-478 Warsaw, Poland\\
$^{2}$Department of Astrophysics/IMAPP, Radboud University, P.O. Box 9010, 6500 GL Nijmegen, The Netherlands\\
$^{3}$Scientific Support Office, Directorate of Science, European Space Research and Technology Centre (ESA/ESTEC),\\ ~Keplerlaan 1, 2201AZ, Noordwijk, The Netherlands\\
$^{4}$Nicolaus Copernicus Astronomical Centre, Polish Academy of Sciences, ul. Bartycka 18, PL-00-716 Warsaw, Poland
}

\date{Accepted XXX. Received YYY; in original form ZZZ}

\pubyear{2017}

\begin{document}
\label{firstpage}
\pagerange{\pageref{firstpage}--\pageref{lastpage}}
\maketitle

\begin{abstract}
We investigate the impact of combining Gaia astrometry from space with precise, high cadence OGLE photometry from the ground.
For the archival event OGLE3-ULENS-PAR-02, which is likely a black hole, we simulate a realistic astrometric time-series of Gaia measurements and combine it with the real photometric data collected by the OGLE project. 
We predict that at the end of the nominal 5 years of the Gaia mission, for the events brighter than $G\approx 15.5$ mag at the baseline, caused by objects heavier than 10 \msun, it will be possible to unambiguously derive masses of the lenses, with accuracy between a few to 15 per cent. We find that fainter events ($G<17.5$) can still have their lens masses determined, provided that they are heavier than 30 \msun. We estimate that the rate of astrometric microlensing events caused by the stellar-origin black holes is $\approx 4 \times 10^{-7} \, \rm yr^{-1}$, which implies, that after 5 years of Gaia operation and $\approx 5 \times 10^6$ bright sources in Gaia, it will be possible to identify few such events in the Gaia final catalogues.
\end{abstract}




\section{Introduction}
Mass is one of the most important parameters of a star as it determines its
entire evolution. Even though astronomers are able to derive many properties of stars
using various observing techniques (e.g. \citealt{2010ARA&A..48..581S}, \citealt{2012ApJ...757..112B}), mass has always been the problematic one.
It is possible to estimate mass from spectral classification, but precise determination can be only achieved in
very specific circumstances, for example double-lined eclipsing binaries (see \citealt{2010A&ARv..18...67T} for a review).

Detecting and measuring the mass of neutron stars (NS) or stellar mass black holes (BH) is even more challenging,
as they usually do not emit any light.
We can observe gamma ray emission or X-ray emission from very young neutron stars (e.g. \citealt{2007Ap&SS.308..181H}, \citealt{2010ApJ...725L..73A}) or radio signal from pulsars \citep{2008A&A...482..617P}. 
Measuring the timing residuals of the pulsations can yield very accurate mass measurements for NSs (\citealt{1977puls.book.....M}, \citealt{2013ApJ...778...66K}).
Another example is the class of objects known as the X-ray binaries, where a BH or a NS can feed itself with the matter from a companion \citep{2010ApJ...725.1918O}. 
Moreover, the recent discovery of gravitational waves \cite{2016PhRvL.116f1102A} with the Laser Interferometer Gravitational-Wave Observatory (LIGO, \citealt{2009RPPh...72g6901A}) shows that it is possible to observe a merging system containing two BHs, and to derive physical information about the components and the final merged object, including their masses.

Every example mentioned above (besides the rare case of the young neutron stars) requires a
BH or NS to be in a binary system.
Characteristics of compact objects in such systems may be different than single objects, because they have gone through different evolution.
It is estimated that about $10^8$ stellar mass black holes reside in our Galaxy (\citealt{2000ApJ...535..928G}, \citealt{2002MNRAS.334..553A}), but there is no confirmed detection of the single BH yet. Thus, detecting and measuring the masses of such objects would strongly enhance our understanding of the origin, evolution and distribution of BHs in our Galaxy.

This is where the gravitational microlensing method reveals its advantages. 
When considering a single BH as a lens, problem of its darkness is non-existent, as the only light needed comes from the more distant source.
The formula for the lens mass (e.g. \citealt{2000ApJ...542..785G}) reads:
\begin{equation}
M = \frac{\theta_E}{\kappa \pi_E} = \frac{t_E \mu_{rel}}{\kappa \pi_E}~, ~~~ \pi_E = \frac{\pi_{rel}}{\theta_E}~, ~~~ \pi_{rel} \equiv \frac{1}{D_l}-\frac{1}{D_s}
\end{equation}
where $D_l$ and $D_s$ are distances to the lens and the source respectively, $t_E$ is the timescale of the event, $\mu_{rel}$ is the relative proper motion between the lens and the source, $\kappa\approx 8.14 \frac{mas}{M_{\odot}}$ is constant, $\theta_E$ is the angular Einstein radius and $\pi_E$ is the microlensing parallax, which reflects the influence of the Earth orbital motion onto the course of an event (\citealt{2000ApJ...542..785G}, \citealt{2004ApJ...606..319G}). 
Both $\theta_E$ and $\pi_E$ are measurable, although both require particular circumstances to occur. 

To measure the microlensing parallax $\pi_E$, the event has to be observed from different positions in the solar system. One way to achieve this is to observe over a long period of time, long enough for the Earth to move noticeably on its orbit (e.g. \citealt{2002ApJ...579..639B}, \citealt{2005ApJ...633..914P}). This means that mostly events with longer timescales $t_E$ can reveal a significant microlensing parallax signal. Another method is to observe the same event simultaneously from the ground and from a space satellite, as the difference between the two lightcurves obtained from two different points of view allows us to derive the microlensing parallax signal (e.g. \citealt{1966MNRAS.134..315R}, \citealt{1995ApJ...441L..21G}). Such measurements are being conducted with the Spitzer satellite (e.g. \citealt{2015ApJ...804...20C}, \citealt{2015ApJ...799..237U}) and may also be possible for some microlenses detected by Gaia, for example the spectacular binary event Gaia16aye \citep{2016ATel.9507....1W}.

The Einstein radius $\theta_E$, which provides a natural scale for a microlensing event, can be derived from the finite source effect (\citealt{1986ApJ...304....1P}, \citealt{1997ApJ...491..436A}). The idea is to retrieve the size of the source in $\theta_E$ units during the caustic crossing in binary lens event or in the high magnification event (e.g. \citealt{2012ApJ...751...41C}). With additional information about the angular size of the source one can determine the Einstein radius. Unfortunately, such high magnification cases are rare and information about the source size is not always certain.

Since $\theta_E=t_E \mu_{rel}$, and the timescale $t_E$ can be derived from the lightcurve,
measuring the relative proper motion of the lens and the source is another way to determine $\theta_E$.
If one waits long enough for the lens and source to separate, the measurement of the proper motion will be straightforward. Taking into account that the relative proper motion is of the order of few miliarcseconds per year, it would take many years. Moreover, if the lens is a dark stellar remnant, this approach is obviously not viable.

We will focus on yet another way of obtaining $\theta_E$, making use of the fact that microlensing not only changes the brightness of the source, but also its apparent position (\citealt{1995ApJ...453...37W}, \citealt{1995A&A...294..287H}, \citealt{1995AJ....110.1427M}). 
As two images emerge during the event, they are both unresolvable and of different brightness. It leads
to the shift of the centroid of light from the original source position towards the brighter image. This phenomena is referred to as astrometric microlensing \citep{Dominik2000}. The centroid shift scales with the Einstein radius $\theta_E$. Thus, for events yielding detectable microlensing parallax signal, observing both photometric and astrometric parts of the microlensing will enable the measurement of $\pi_E$ and $\theta_E$, which is sufficient to determine the mass of the lens.

The positional change of the centroid depends on the $\theta_E$ and separation $u$. Contrary to the photometric case, the maximum shift occurs at $u_0=\sqrt{2}$ and reads \citep{Dominik2000}
\begin{equation}
\delta_{max}=\frac{\sqrt{2}}{4} \theta_E \approx 0.354 \theta_E.
\end{equation}
Thus, for the relatively nearby lens at $D_l = 4kpc$, source in the bulge $D_s=8kpc$ and lensing by a stellar BH with the mass $M=4M_{\odot}$, the astrometric shift due to microlensing will be about 0.7 milliarcsecond. Such precision is hard to obtain in current ground-based and space-based observatories. Nonetheless, there are approaches to detect this subtle effect for some extreme cases. It was possible to give wide constraints on the mass of the lens in three known microlensing events by having a non-detection of the centroid shift using Keck adaptive optics \citep{2016ApJ...830...41L}. The most spectacular result in this field to date is recent mass measurement for the very nearby white dwarf, thanks to the first ever detection of the astrometric microlensing signal with the Hubble Space Telescope \citep{2017Sci...356.1046S}.

Another great opportunity to detect the astrometric shift during a microlensing event may come with the Gaia mission \citep{2016A&A...595A...1G}.
The  satellite was launched in December 2013 and it began full operations in July 2014. The first catalogue of the average positions of more than a billion stars was released in September 2016 \citep{GaiaDR1}. 
While Gaia will primarily provide measurements of proper motions and parallaxes of a billion stars, in its final data release in around 2022, it will also provide astrometric time-series for all stars, which can be used for astrometric microlensing measurements for all detected events. 


In this work we investigate if the astrometric measurements provided by the Gaia mission, combined with very accurate, high cadence photometry from the OGLE project \citep{2015AcA....65....1U} will be sufficient to determine the masses of lenses in microlensing events towards the Galactic bulge, and especially to distinguish between stellar lenses, and events resulting from dark compact massive objects. 
In this paper we present the analysis of the archival event OGLE-2006-BLG-095 found by EWS \citep{1994AcA....44..227U}, which is likely due to a BH \citep{2016MNRAS.458.3012W}. 
Adopting the nomenclature from Wyrzykowski et al., the event will be referred to as OGLE3-ULENS-PAR-02, hereafter PAR-02. We want to recognise the possibilities arising from the combination of Gaia and OGLE measurements considering only this particular event as an example. We discuss few examples of possible astrometric trajectories which could explain the observed PAR-02 lightcurve or similar events of different baseline brightness. Knowing predictions for the astrometric precision for the Gaia mission, scan directions and epoch for every measurement, we simulate realistic Gaia {\it per epoch} astrometry for these trajectories.
We derive and present a method for combining the astrometric and photometric data sets in anticipation of future availability.


We conclude that combining precise astrometry provided by the Gaia mission and photometry from ground based surveys like the OGLE-IV project for microlensing events, it will be possible to derive the masses of the lenses in the brightest events, with accuracies in the range of a few to 15 per cent, depending on the mass and brightness of the event, and therefore to judge whether the lens is a BH or not.

The paper is organised as follows. 
We first describe the astrometric and photometric model for a single microlensing event and then we present a framework for the simulation of the Gaia astrometric time series.
We then analyse the combined real photometric data for OGLE3-ULENS-PAR-02 event with mock astrometric data and derive the parameters of the lens. 
We discuss the results in Section 4. Section 5 is dedicated to a discussion on rates and detectability of black hole lenses. We conclude in Section 6.

\section{Astrometric and photometric microlensing model} 
There are many complete and interesting reviews for both photometric
(\eg \citealt{1964MNRAS.128..295R}, \citealt{1986ApJ...304....1P}, \citealt{2006glsw.conf..453W}) and astrometric
flavours of microlensing (\eg \citealt{1995ApJ...453...37W}, \citealt{1998ApJ...502..538B}, \citealt{Dominik2000}).
Some authors have considered combining these two instances of microlensing
as well (\eg \citealt{1995AJ....110.1427M}, \citealt{2002MNRAS.331..649B},
\citealt{2016ApJ...830...41L}). 
In particular Belokurov and Evans investigated how Gaia can help measuring microlensing parameters of lenses, however, only now, the Gaia performance parameters are known well enough and Gaia astrometry can be reliably simulated. We recall their work in more details in Section 4.

\subsection{Photometric model with parallax}
\label{sec:photoModel}
The most simplistic approach to the microlensing phenomena is to assume point mass lens model,
point source and their linear relative motion. 
Such model consists of five parameters: the time of maximum amplification $t_0$,
the projected separation $u_0$ between the source and the lens at the time $t_0$, blending parameter $f_s$,
the baseline brightness $I_0$ and timescale of the event $t_E$. The formula for amplification reads (e.g. \cite{1986ApJ...304....1P}):
\begin{equation}
A(t) =  \frac{u^2(t) + 2}{u(t)\sqrt{u^2(t) + 4}},
\end{equation}
where $u(t)$ is the projected separation of the source and the lens at the time $t$:
\begin{equation}
u(t)=\sqrt{u_0^2 + \tau_E^2(t)}~, ~~~\tau_E(t)=\frac{t-t_0}{t_E}
\end{equation}
Because the lensing only affects the light from the source, the total flux during the event can be denoted as
\begin{equation}
F_{tot}(t) = A(t) F_s + F_b,
\end{equation}
where $F_s$ is the flux from the source and $F_b$ is the flux from the blend, which can be either from the lens, another star very close to the line of sight or both. 
For convenience the blending parameter is defined as
\begin{equation}
f_s=\frac{F_s}{F_s+F_b},
\end{equation}
which can then be used to rewrite the formula for the total flux during the event as
\begin{equation}
\frac{F_{tot}(t)}{F_0} = A(t) f_s + (1-f_s),
\end{equation}
where $F_0 = F_s + F_b$ is the total flux outside the microlensing event (in the baseline). 
Finally, knowing the equations (2-6), it is possible to construct the formula for the lightcurve model in the standard microlensing event:
\begin{equation}
I(t) = I_0 - 2.5 \log{F_{tot}(t)}
\end{equation}
Nevertheless, the physical quantities like mass of the lens or distances to the source and to the lens can not be
derived from a standard, five-parameter modelled lightcurve. As shown in equation (1), the microlensing parallax is essential (although still
not sufficient) to retrieve the lens mass from the photometric data. For that, our model has to be extended to include the parallax
effect, due to Earth orbital motion. 
Although calculation of the parallactic deviations is not trivial, it can be simply denoted by redefining $u(t)$
components (see \cite{2004ApJ...606..319G}):
\begin{equation}
\tau(t) = \tau_E(t) + \delta\tau~, ~~~\beta(t) = u_0 + \delta\beta
\end{equation}
The $(\delta\tau, \delta\beta)$ correction changes with the orbital motion of Earth and thus make the source-lens projected
separation non-linear in time. 
In practice, the microlensing parallax $\vec{\pi_E}=(\pi_{EN}, \pi_{EE})$ parameter is used in the modelling, 
and $(\delta\tau, \delta\beta)$ are derived afterwards. 
With two additional parameters $\pi_{EN}$ and $\pi_{EE}$ we can define new separation parameter that includes parallax:
\begin{equation}
u_{par}(t) = \sqrt{\beta^2(t) + \tau^2(t)}.
\end{equation}
Making this one change in equations (2-7), we can introduce the seven-parameters ($t_E, u_0, t_0, f_s, I_0, \pi_{EN}, \pi_{EE}$)
microlensing model with parallax.

Figure \ref{fig:photometry} shows the photometry from the OGLE-III survey of an event OGLE3-ULENS-PAR-02 with both standard  five-parameter and seven-parameter parallax models. 

\subsection{Astrometric model}
\label{sec:astroModel}
In a regular single lens case, the two images of the source emerge, but are not separable since their distance is of order of 1 mas.
What can be measured more easily is the shift of the centroid of light of the two images.
The offset in the microlensing event is described, following \cite{Dominik2000}, as:
\begin{equation}
\delta(t) = \frac{\sqrt{u_0^2 + \tau_E^2(t)}}{u_0^2+\tau_E^2(t)+2},
\end{equation}
with components $\delta_{\parallel}$, parallel to the motion of the lens relative to the source and $\delta_{\bot}$ perpendicular to it, both in the direction away from the lens as seen from the source:
\begin{equation}
\begin{array}{c}
\delta_{\parallel} = \frac{\tau_E(t)}{u_0^2 + \tau_E^2(t) + 2}\theta_E, \\ \\
\delta_{\bot} = \frac{u_0}{u_0^2 + \tau_E^2(t) + 2}\theta_E.
\end{array}
\end{equation}
These equations result in an ellipse with semi-major axis $a$ and $b$:
\begin{equation}
a = \frac{1}{2\sqrt{u_0^2 + 2}}\theta_E~, ~~~
b = \frac{u_0}{2(u_0^2 + 2)}\theta_E.
\end{equation}
Note that this ellipse represents the centroid movement relative to the source therefore the full trajectory will be the superposition
of this ellipsoidal motion and the source proper motion. Moreover, its orientation depends on the direction of the $\vec{\mu_{rel}} = \vec{\mu_{lens}} - \vec{\mu_{src}}$ vector. These aspects will be discussed in the next subsection.

Similar to the photometric case, after replacing standard projected separation from equation 3. with the parallactic one
from equation 9. in equations 10-12, we can obtain astrometric deviation occurring during the microlensing event which
includes the parallax effect. It results in wiggles on the astrometric ellipse caused by the Earth orbital motion, as shown in Fig. \ref{fig:ellipse}.

What is important in our case, the astrometric deviation of the centroid
is not decreasing as quickly as the total photometric magnification with the lens-source projected separation.
The formula for magnification (e.g. \citealt{1986ApJ...304....1P}) falls off rapidly with separation, as
\begin{equation}
A = \frac{u^2 + 2}{u\sqrt{u^2 + 4}} \xrightarrow[u \gg 1]{} 1 + \frac{2}{u^4}.
\end{equation}
The case of the astrometric microlensing is more forgiving for sparse but long data, because for
the centroid shift (e.g. \cite{Dominik2000}) we have
\begin{equation}
\delta = \frac{u}{u^2 + 2}\theta_E \xrightarrow[u \gg 1]{} \frac{1}{u} \theta_E.
\end{equation}

\subsection{Complete motion curve model}

\begin{figure}
\begin{center}
\includegraphics[width=\columnwidth]{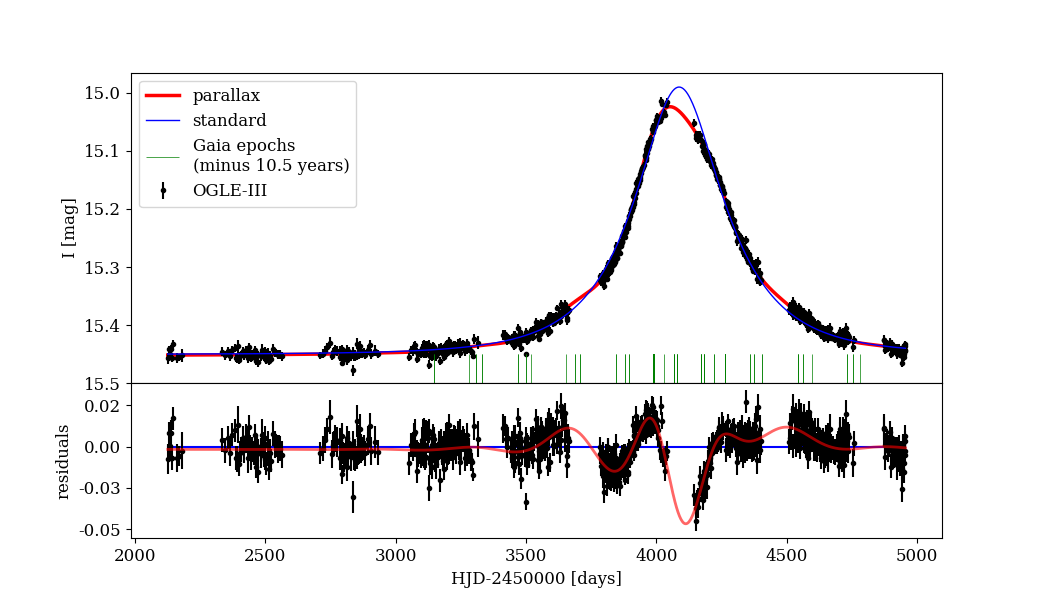}
\caption{Photometric data from multiple seasons of the OGLE-III project of the event OGLE3-ULENS-PAR-02. 
Standard and parallax model are shown with relevant residuals below. The green lines mark Gaia epochs for the coordinates of PAR-02 as if the event took place during the lifetime of the mission.
}
\label{fig:photometry}
\end{center}
\end{figure}

\begin{figure}
\begin{center}
\includegraphics[width=\columnwidth]{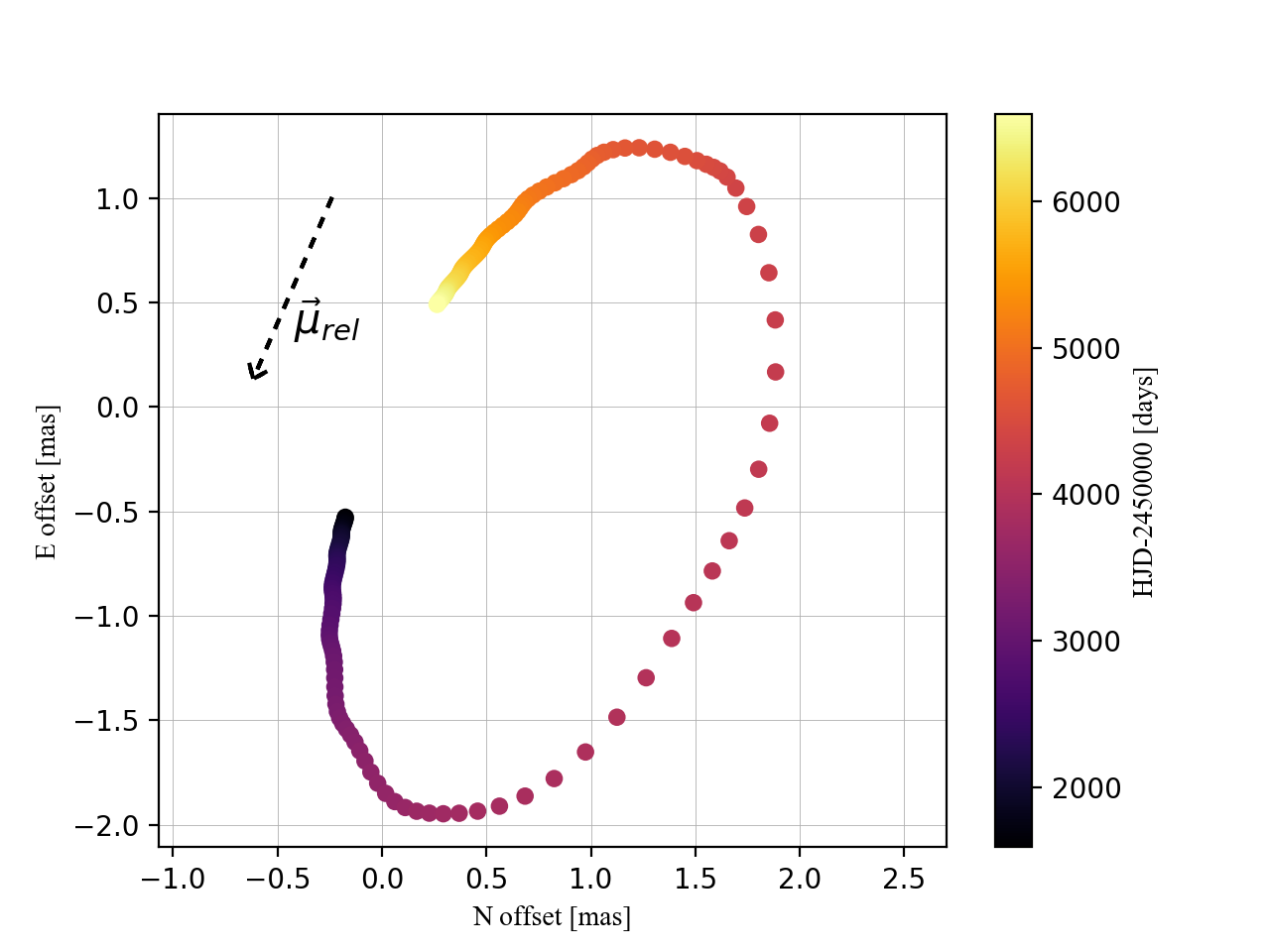}
\caption{Trajectory of the centroid of light for the case of PAR-02 as seen relative to the source, which is located at (0,0) here. The deviation from the ellipse shape is due to the parallax effect. As $t_0 \approx 4100~days$,  even long before (and after) the event, the centroid is noticeably shifted from the original source position.}
\label{fig:ellipse}
\end{center}
\end{figure}

To align the ellipsoidal trajectory of the centroid relative to the source position (Fig. \ref{fig:ellipse}), we need the direction of the vector $\vec{\mu_{rel}}$. 
Its direction is related to the orientation of the ellipse, while its length impacts the absolute amplitude of
the deviation (as both are proportional to $\theta_E$). To reproduce the motion of the centroid
as seen by an observer on Earth (or by Gaia), we need to add the motion of the source, which is represented by the
$\vec{\mu}_{src}$ vector and account for the orbital motion of the observer. 
We can write the formula for the position of the centroid on the sky as:
\begin{equation}
\vec{\xi}(t) = \vec{\xi}_{0} + \vec{\mu}_{src}t + \vec{\delta}(t)+\vec{\Pi}_{src}(t), 
\end{equation}
where $\vec{\xi}_{0}$ is a reference position of the source, $\vec{\delta}$ is the position of the centroid relative to the source (Eq. 10.) and $\vec{\Pi}_{src}$ is the change
of the source trajectory due to the parallax effect, and $\Pi_{src} = \frac{1}{D_s}$. 
The examples of four centroid trajectories are presented on Figure \ref{fig:fourclasses}.

To calculate and include the parallax of the source, we need to know the distance to it. We assume that
the source resides in the Galactic bulge and $D_s = 8~$kpc. This assumption is justified by the
fact that most of the sources are giants from the bulge old population (Red Clump Stars). 
This is at least true in case of our studied event, PAR-02.
Moreover, we aim for events with high relative parallax $\pi_{rel}$ and majority of such events will have a distance source lying in the bulge and nearby lens from the disk. In case of the events from the Galactic disk, it might be necessary to infer the source distance from the spectroscopic measurements.

\subsection{Astrometry in the Gaia mission}
\label{sec:gaia}

Gaia is designed, in principle, to derive parameters like parallax and proper motions for a billion of stars from our part of the Galaxy.
It will utilise astrometric time-series collected over five years of the mission in a global astrometric solution (\citealt{2012A&A...538A..78L}, \citealt{2016A&A...595A...4L}), meaning that all parameters for all stars will be derived simultaneously in a huge equation. 

However, an astrometric anomaly due to a microlensing event will complicate the solution and the individual time-series will have to be retrieved for known microlensing events discovered during the duration of the mission. 

 
 \subsubsection{Gaia's scanning law}
 
\begin{figure*}
\begin{center}
\includegraphics[width=10cm]{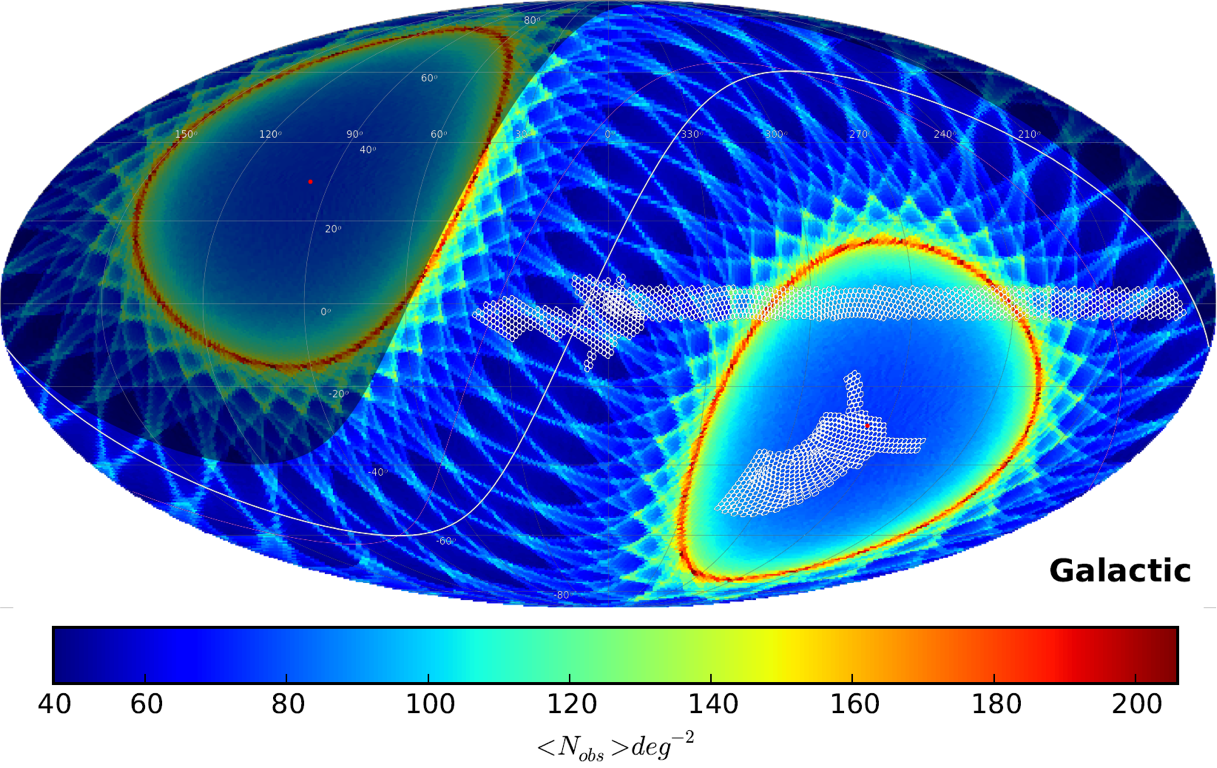}
\caption{Number of transits as estimated for the end of the nominal 5 years mission, presented in Galactic coordinates and OGLE-IV footprint. Image
credit Jan Skowron and Nadejda Blagorodnova.}
\label{fig:scanning}
\end{center}
\end{figure*}

 Gaia operates and registers signal from objects in a very specific way, which makes the cadence and produced output
 somewhat unusual (see Appendix A in \citealt{2017arXiv170203295E}). 
 It uses a Lissajous orbit around the L2 point and revolves around its own axis, which is perpendicular
 to the line of sight of its two telescopes (mirrors). The combination of these facts, and additional precessed motion of the axis
 allows Gaia to cover the whole sky, producing a peculiar pattern of observations density, as shown in Figure \ref{fig:scanning}. This figure also shows OGLE-IV footprint covering Bulge, Disk and the Magellanic System \cite{2015AcA....65....1U}.

 On average, Gaia will gather few data points, for every object it detects, once in about 30 days. The most observations
 will be gathered around the ecliptic latitudes $+/- 45^o$ due to the fixed axis between the Gaia's spin axis and the Sun. 
 In this region there will be more than 200 transits, while in poorly covered areas, only about 40 transits are expected.
 The Galactic bulge and disk will not be observed more than 100 times during the 5 years mission, apart from a small section of the disk in both North and South parts at $+/- 45^o$ ecliptic latitude. 
 Such cadence makes it not optimal for the photometric microlensing measurements. 
 It usually will not be enough to accurately cover the lightcurve, even for longer timescale events.
 In case of PAR-02 the coverage is above average, because it lies in the relatively densely probed region of the bulge - 92 measurements during the five years of the mission, which is almost two times more points than the average for bulge.
 However, as mentioned in subsection \ref{sec:astroModel},
 astrometric deviation is detectable long before and after the time of the maximum magnification 
 (see Figure \ref{fig:fourclasses}), which makes astrometric microlensing a well suited target for the Gaia mission.
 
 As the scanning law of Gaia is predefined and well known, it is possible to estimate the time when the spacecraft will observe at given coordinates. 
 The GOFT (Gaia Observing Forecast Tool, https://gaia.esac.esa.int/gost/) allows to do so, although
the predicted and actual time of observation may vary as some effects are not taken into account in the calculations.

\begin{table*}
\begin{center}
\begin{tabular}{|c|c|c|c|c|c|c|c|c|c|}
\hline
V-I[$mag$]&12&13&14&15&16&17&18&19&20 \\
\hline
0.97& 47& 43& 72& 116& 192& 330& 604& 1193& 2603\\
1.22& 47& 46& 68& 110& 181& 310& 562& 1106& 2380\\
1.77& 28& 42& 59& 95& 156& 264& 470& 903& 1892\\
2.17& 29& 43& 53& 86& 141& 236& 414& 782& 1609\\
2.53& 26& 42& 47& 76& 123& 205& 357& 662& 1334\\
2.77& 31& 44& 47& 69& 112& 185& 320& 586& 1161\\
2.97& 28& 30& 45& 64& 104& 171& 292& 528& 1033\\
3.20& 25& 28& 41& 59& 94& 155& 262& 469& 902\\
\hline
\end{tabular}
\end{center}
\caption{The modified CCD-level centroid location errors in microarcseconds, as a function of Johnsons V magnitude for one focal plane transit, in the AL direction. The centroiding errors were increased by 50\% to account for the calibration errors. Then we lower the error by a factor of $\sqrt{9}=3$ to include measurements from multiple AFs. As a result, the numbers presented here are centroiding errors divided by 2.}
\label{tab:centroidingAL}
\end{table*}

\begin{table*}
\begin{center}
\begin{tabular}{|c|c|c|c|c|c|c|c|c|c|}
\hline
V-I[$mag$]&12&13&14&15&16&17&18&19&20 \\
\hline
0.97& 636& 1010& 3118& 4912& 8492& 16640& 36738& 87530& 177856\\
1.22& 602& 958& 2926& 4590& 7896& 15324& 33414& 79154& 164262\\
1.77& 524& 832& 2530& 3888& 6552& 12318& 26142& 61098& 133560\\
2.17& 472& 748& 2258& 3436& 5700& 10470& 21776& 49976& 114344\\
2.53& 418& 660& 1058& 3010& 4912& 8802& 17750& 39732& 94894\\
2.77& 384& 604& 966& 2748& 4416& 7766& 15338& 33794& 80326\\
2.97& 356& 560& 894& 2544& 4044& 7016& 13588& 29480& 69570\\
3.20& 326& 512& 816& 2336& 3650& 6244& 11866& 25192& 58684\\
\hline
\end{tabular}
\end{center}
\caption{The modified CCD-level centroid location errors in microarcseconds, as a function of Johnsons V magnitude for the SM chip in the AC direction. To obtain numbers in this table we increased centroiding errors by the factor of 2, to account for the calibration errors (see text for more details).}
\label{tab:centroidingSM}
\end{table*}

\begin{figure}
\begin{center}
\includegraphics[width=\columnwidth]{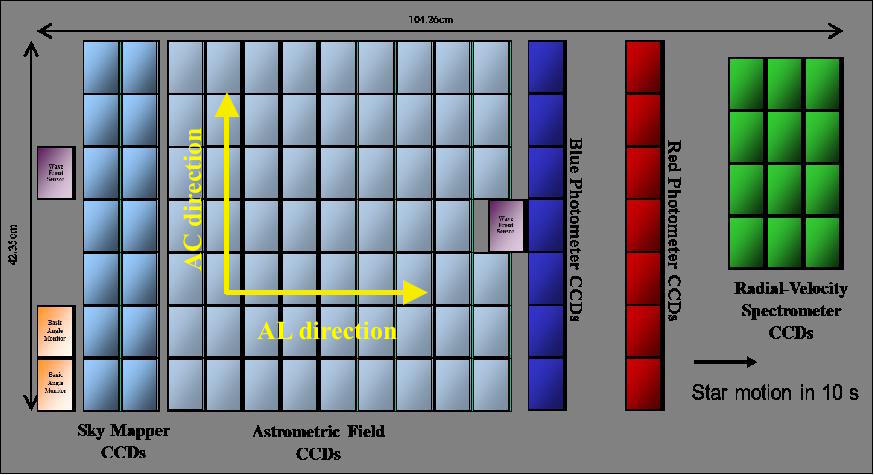}
\caption{Schematic picture of the Gaia focal plane instruments (see text for description). Image credits: ESA.}
\label{fig:CCD}
\end{center}
\end{figure}

\subsubsection{Objects detection and measurements}
 
 Gaia operates using two mirrors and only one focal plane, where the light is received by the CCD camera (Figure \ref{fig:CCD}).
 The camera consists of the sky mapper chips (SM), astrometric fields (AF), low resolution spectrometers in blue and red parts
 of the spectrum (BP and RP respectively) and high resolution spectrometer for radial velocities (RVS). 
 When, due to constant rotation
 of the spacecraft, an object appears in the field of view of one of the telescopes, the window is assigned to it on the SM1 or
 SM2 chip, depending which telescope had detected the object. Then, its position and brightness is measured on
 astrometric fields from AF1 to AF9. After that, the low and high resolution spectra are taken to complete the transit of the object
 through the focal plane. The whole passage lasts for about 45 seconds. The pixels of AF chips are highly elongated in the direction
 across the scan and the position on the chip is derived from the time the object was detected on a given pixel.
 It results in a significant difference between the precision of a position measurements in AL (along scan) and AC (across scan)
 directions, with the AL-direction accuracy much higher than the AC. The spin orientation of Gaia constantly changes so do AL and AC directions. It means that the orientation of the error-bars of astrometric measurements will vary as well, depending on the time the observation
 was taken. For more detailed description of the Gaia observing routine see \citealt{2016A&A...595A...1G}.
 
 The astrometric accuracy will naturally differ with magnitude and colour of the object.
 We estimate Gaia's astrometric accuracy using realistic Monte-Carlo centroiding simulations. These simulations generate objects transiting a CCD and then determine the standard error with which the centroid of the transiting source can be determined. The centroiding simulations properly include a variety of star colours, the variations of the optical quality over the field of view, which affect the sharpness of the Linear Spread Function (LSF, see \citealt{2016A&A...595A...1G} for a detailed explanation). They also include sky background (and associated noise), photon noise and readout noise of the detector and also the finite window size, which is responsible for flux losses. The LSFs are realistic and include all "smearing effects" such as attitude (rate) errors, radiation damage, charge diffusion, pixel binning, optical aberrations, etc. The centroiding errors hence "only" represent the detector-level location-estimation error of an isolated source superimposed on a uniform sky background.
 Table \ref{tab:centroidingAL} shows the expected error-bar at one AF CCD for a range of colours and magnitudes. 
 For AL direction we arbitrarily add 50 per cent to the centroiding error to account for various calibration errors. On the other hand, we utilise the fact that there are nine AFs and thus the accuracy can be improved by a factor of $\sqrt{9} = 3$. As a result, we adopt astrometric error in the AL direction as about two times smaller than the centroiding errors. We present final values for AL-direction for different magnitudes and colours in Table \ref{tab:centroidingAL}.
 
\begin{figure}
\begin{center}
\includegraphics[width=\columnwidth]{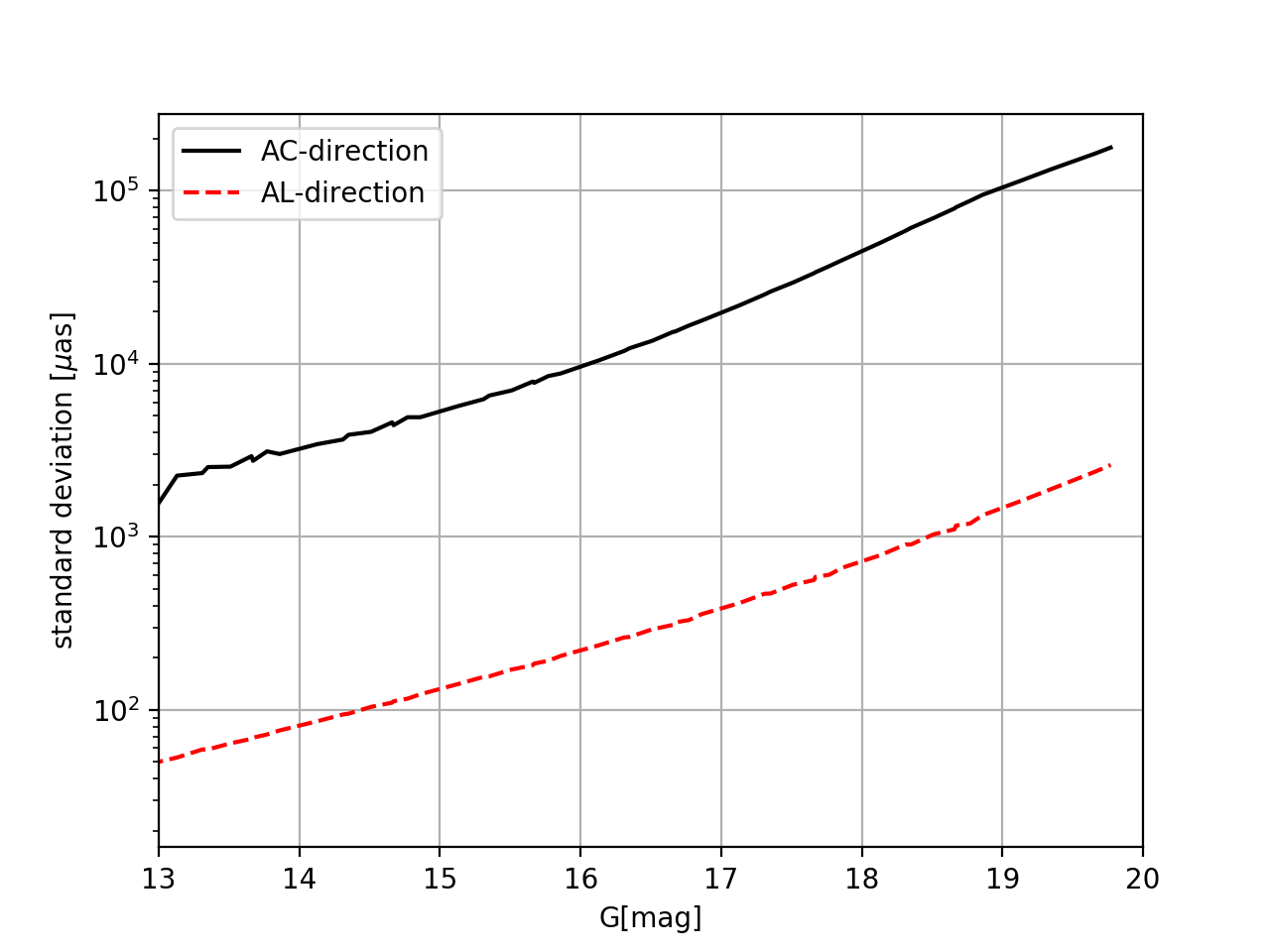}
\caption{The error in the single-transit position measurements for AL-direction (red dashed line) and AC-direction (black solid line) as a function of G-magnitude of the object. 
The accuracy in AC direction is based on SM CCDs, while in the AL is based on all AF CCDs. See text for details. 
For measurements brighter than $G=13$ mag, the 2-D data is collected from the AFs and a different approach has to be applied.}
\label{fig:Gband_sigma}
\end{center}
\end{figure}

 The case of AC-direction is much less optimistic. First of all, most of the stars ($G$>13 mag) will only have position measurements in the AL-direction from the AFs. Thus we utilise the SM CCD measurement for the AC direction errors. It provides lower centroiding accuracy, moreover, there is only one SM measurement per transit so we do not benefit from the multiple measurements per transit as we did for the AL-direction. Also, the calibration errors for the SM chips will be higher - we adopt 100\% here (more on that in the next paragraph). Thus, the AC-direction error used in our simulations is two times higher than the centroiding error for the SM chip. The final values for AC-direction for different magnitudes and colours are shown in Table \ref{tab:centroidingSM}. It is worth noting that the colour-index dependency can be removed in Table \ref{tab:centroidingAL} and Table \ref{tab:centroidingSM} when using G-magnitude instead of V-magnitude. On Fig. \ref{fig:Gband_sigma} we present the relation between the G-magnitude and error in both AL and AC directions. In Table \ref{tab:astrom_err} we present the final uncertainties used in simulations for the PAR-02 event ($V-I=2.1$) for a range of magnitudes.
 
 There are three types of errors that we assume to be covered within mentioned 50\% margin for AL and 100\% for AC directions. The centroiding focal-plane locations need to be "projected" onto the sky and subsequently need to be processed in an astrometric model. This involves the geometric calibration of the Focal Plane Array (which includes issues such as chromaticity corrections, column-response non-uniformity, Prototype Engineering Module non-uniformity, Charge Transfer Inefficiency corrections, basic-angle variations, etc., including temporal variations of all these) and the attitude reconstruction of the spacecraft (which includes issues such as micro-meteoroids, micro-clanks, etc.).
Secondly, there are scientific uncertainties and residual calibration errors in the on-ground data processing and analysis, for example uncertainties related to the spacecraft and solar-system ephemerides, estimation errors in the sky background value that needs to be fed to the centroiding algorithm, the contribution to the astrometric error budget resulting from the mismatch between the actual source LSF and the calibrating LSF ("template mismatch"), etc. Finally, there are environmental, modelling "errors": the centroiding simulations assume a perfect, photometrically non-variable star, in isolation, with a uniform sky background. The real sky, on the other hand, has crowding, sky-background inhomogeneities, etc. To account for all these we introduce 50\% margin for AL and 100\% for AC directions mentioned earlier.
The reason for a larger error in AC
direction is that the SM CCDs have a poorer
geometric calibration than the AF CCDs and
that, in general, AC calibrations are worse
than the associated AL versions since the
spacecraft design has been optimised to
deliver optimum AL centroiding performance.

\begin{table}
\begin{center}
\begin{tabular}{|c|c|c|c|c|}
\hline
I-band&G-band&V-band&$\sigma_{AL}$&$\sigma_{AC}$ \\
$[mag]$&$[mag]$&$[mag]$&$[mas]$&$[mas]$ \\
\hline
12 & 13.3 & 14.2 & 0.056 & 2.31 \\
\hline
13 & 14.3 & 15.2 & 0.089 & 3.54 \\
\hline
14 & 15.3 & 16.2 & 0.145 & 5.91 \\
\hline
15 & 16.3 & 17.2 & 0.244 & 10.97 \\
\hline
16 & 17.3 & 18.2 & 0.430 & 23.03 \\
\hline
17 & 18.3 & 19.2 & 0.818 & 52.84 \\
\hline
\end{tabular}
\end{center}
\caption{Standard deviation in the AL and AC directions for different $I$-band and $G$-band magnitudes for PAR-02 event. These are values we used in our simulations as error bars for mock Gaia data. Note that Gaia astrometric accuracy depends on the colour of the object as well. Values presented here are derived for $V-I$=2.1 which is the colour index of the source in PAR-02 event.}
\label{tab:astrom_err}
\end{table}

\section{Case study of OGLE3-ULENS-PAR-02}

\cite{2016MNRAS.458.3012W} searched through the OGLE-III database (2001-2009) for microlensing events with strong parallax signal to find best candidates for lensing BHs. 
They found 13 events for which, in a purely statistical manner, the probability of being a BH (or other dark remnant) is significant, and PAR-02 was the most promising one. 
The authors calculated that its most plausible mass of the lens was $M\sim8.6~M_{\odot}$ which, confined with lack of light from the lens, makes it a strong candidate for a BH event.
We use the original OGLE photometry of PAR-02 and mock Gaia astrometry for extensive modelling and analysis.
Naturally, the event PAR-02 occurred about 10 years before Gaia was launched, therefore all mock Gaia data we generate are shifted in time by about 10.5 years backward. 
Apart from this small tweak, we use real Gaia sampling and its characteristics as they are currently in place. 
However, we use PAR-02 as an example of BH lens events to be discovered during the duration of the Gaia mission either by OGLE-IV survey (in operation since 2010, see \citealt{2015AcA....65....1U}) or by Gaia itself, but then intensively followed-up photometrically from the ground to assure good photometric coverage and good measurement of the parallax signal. 

First step in analysing the PAR-02 event was to obtain initial information about it from
the photometry. While in the real-life situation we would have both astrometric and photometric measurements, in this experiment we only have OGLE photometry and we need to generate the synthetic astrometric signal. Using the model described in section \ref{sec:photoModel} we searched for the best solutions using an implementation of MCMC
(Monte Carlo Markov Chain) by \citet{2013PASP..125..306F}. Parameters from the photometric microlensing will be used to create
possible astrometric trajectory of the centroid. The data and the parallax model for the PAR-02 are shown in Fig. \ref{fig:photometry}.

The microlensing phenomena is a subject to many degeneracies (\eg \citealt{2004ApJ...606..319G}, \citealt{2011ApJ...738...87S}) and in the modelling of the single point mass lens events with the parallax effect one can obtain up to four solutions, all with different parameters, in particular with different microlensing parallax components $\pi_{EN}$, $\pi_{EE}$ and the time-scale $t_E$, affecting mass measurement. 
In case of the photometric model of PAR-02 there were two concurrent solution found in \cite{2016MNRAS.458.3012W}, which we also recover here, and list in Table \ref{tab:model}.

\begin{table*}
\begin{center}
\begin{tabular}{|c|c|c|c|c|c|c|c|}
\hline
solution No.&$t_0$&$t_E^{helio}$&$u_0$&$\pi_{EN}$&$\pi_{EE}$&$f_s$&$I_0$ \\
$[-]$&$[days]$&$[days]$&$[\theta_E]$&$[-]$&$[-]$&$[-]$&$[mag]$\\
\hline
1. & $4091.98^{+0.32}_{-0.30}$ & $254.50^{+10.80}_{-7.86}$ & $-0.870^{+0.055}_{-0.044}$ & $0.0322^{+0.0012}_{-0.0012}$ & $-0.0742^{+0.0049}_{-0.0041}$ & $1.023^{+0.102}_{-0.117}$ & $15.4522^{+0.0006}_{-0.0006}$ \\
\hline
2. & $4090.72^{+0.26}_{-0.28}$ & $296.52^{+8.22}_{-7.42}$ & $0.664^{+0.027}_{-0.027}$ & $0.0293^{+0.0011}_{-0.0010}$ & $-0.0529^{+0.0026}_{-0.0025}$ & $0.635^{+0.042}_{-0.042}$ & $15.4529^{+0.0006}_{-0.0006}$ \\
\hline
\end{tabular}
\end{center}
\caption{Solutions for photometric data of PAR-02 resulting from the MCMC simulations. We adopt the solution 1 as the real one and use it later to generate mock Gaia measurements. This solution is also slightly preferred in the analysis of this event in \citet{2016MNRAS.458.3012W}.}
\label{tab:model}
\end{table*}

It is in principle not important which set of parameters is used as they are only necessary to create mock Gaia measurements.
With Gaia astrometric data available it will not be needed. For now we have to generate only the example of the astrometric data so we arbitrarily chose first solution (see Table \ref{tab:model}). This solution is also more in favour of a black hole lens, since the blending parameter is nearly 1, indicating very low or no contribution from the lens to the overall light. Lack of blending is also preferable in the case of astrometric microlensing offset, since an additional blending light will diminish the total observed shift. 

\begin{figure*}
\begin{center}
\begin{tabular}{cc}
\includegraphics[width=\columnwidth]{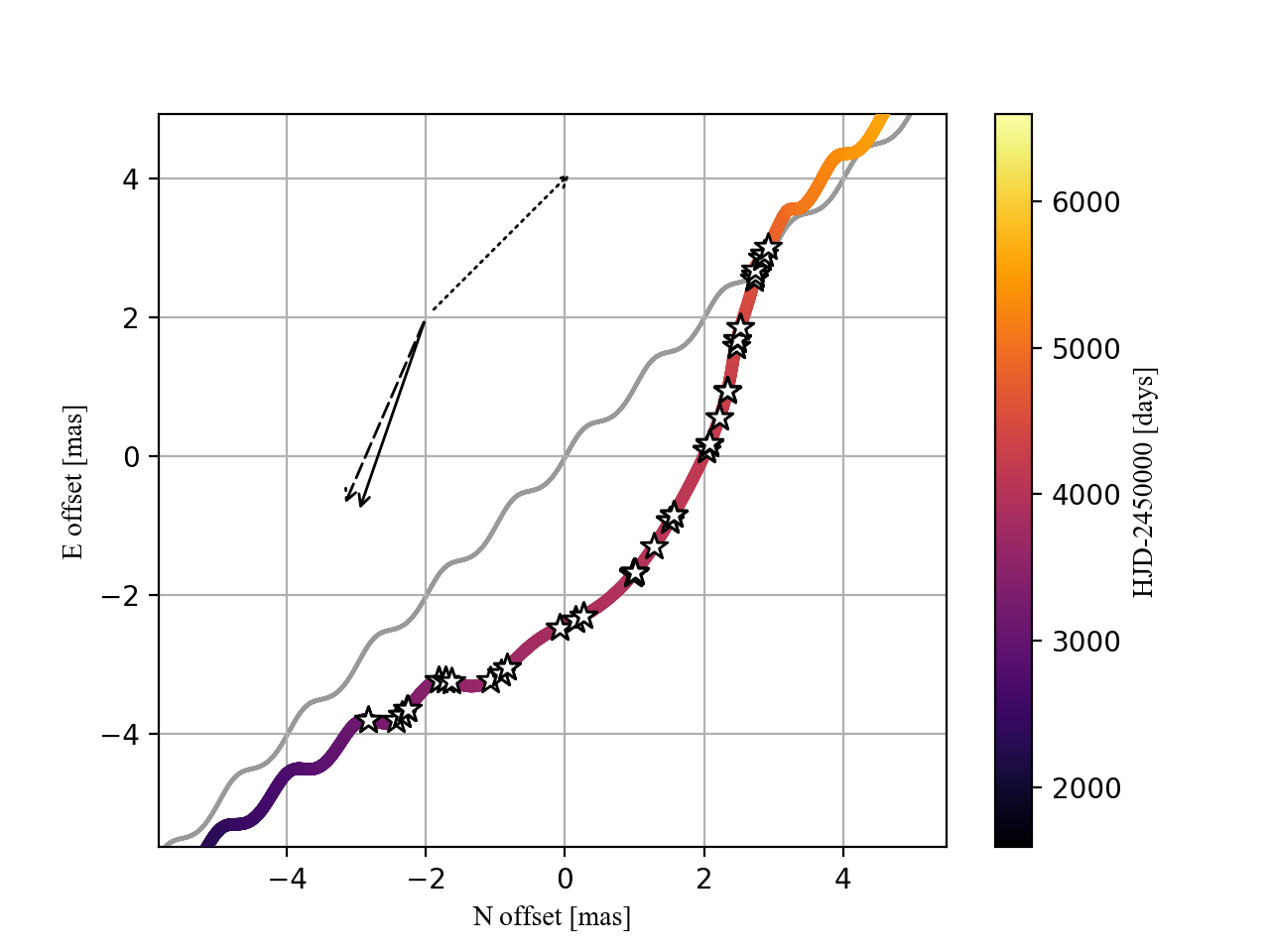}&
\includegraphics[width=\columnwidth]{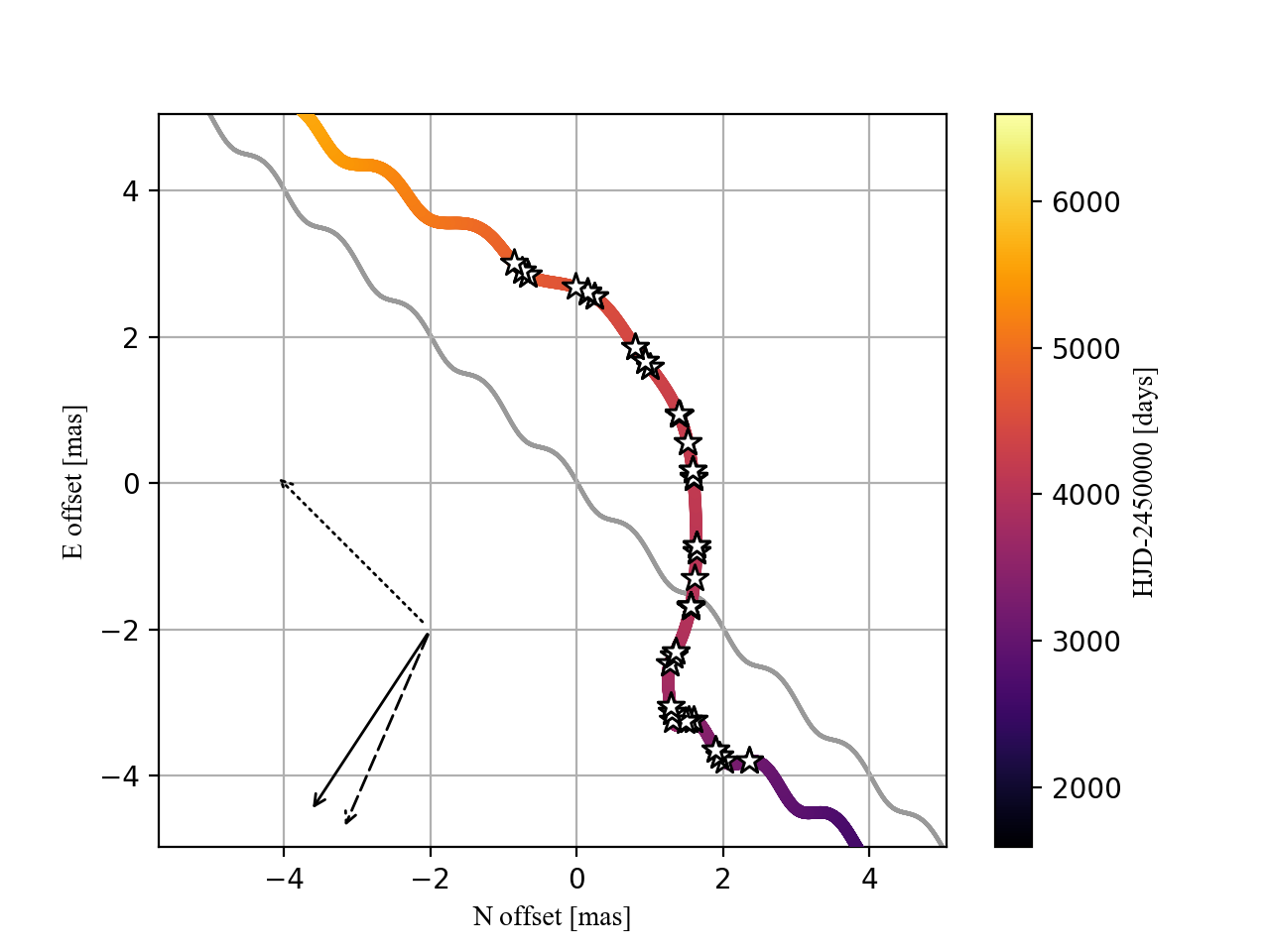}\\
\includegraphics[width=\columnwidth]{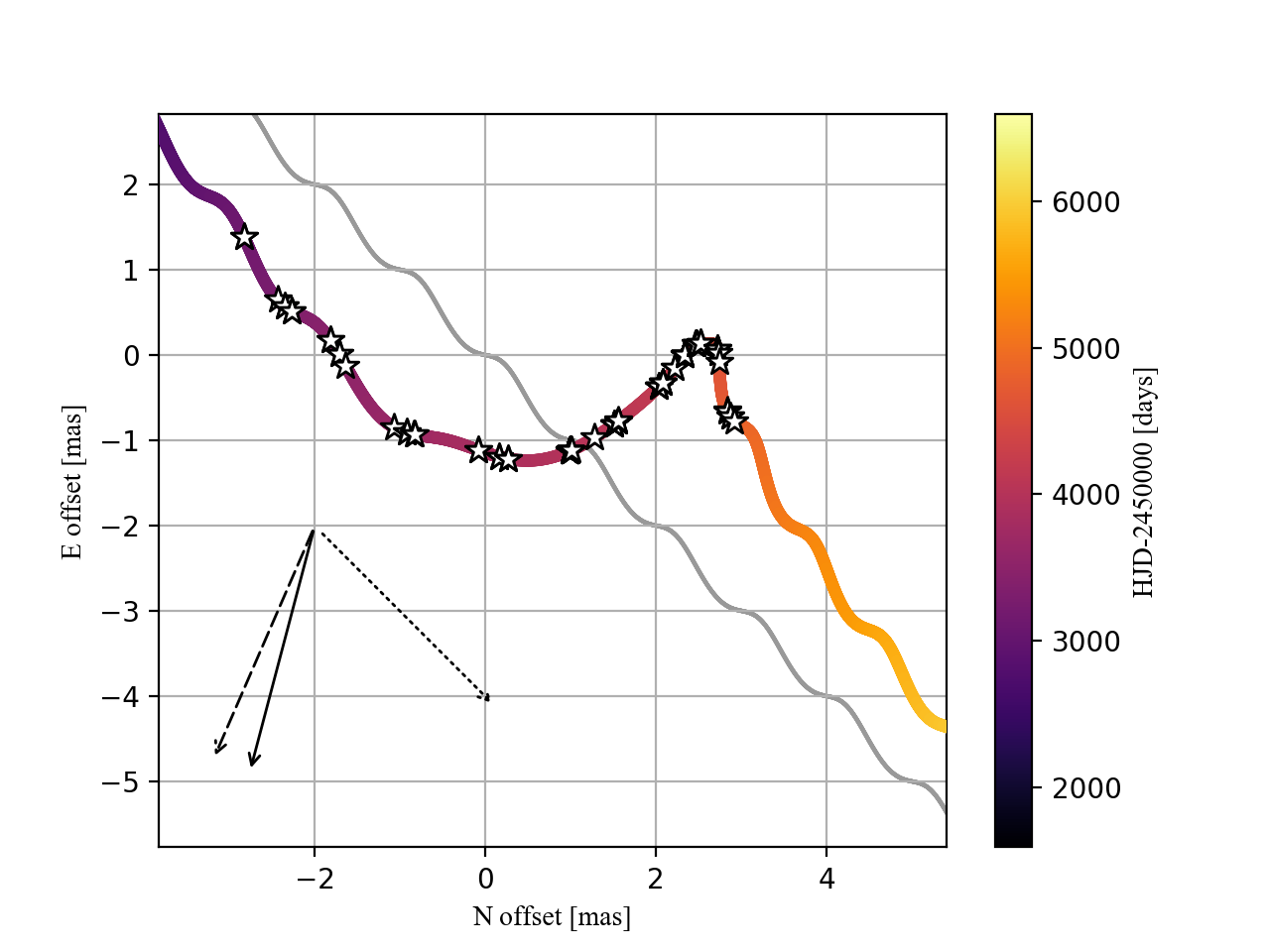}&
\includegraphics[width=\columnwidth]{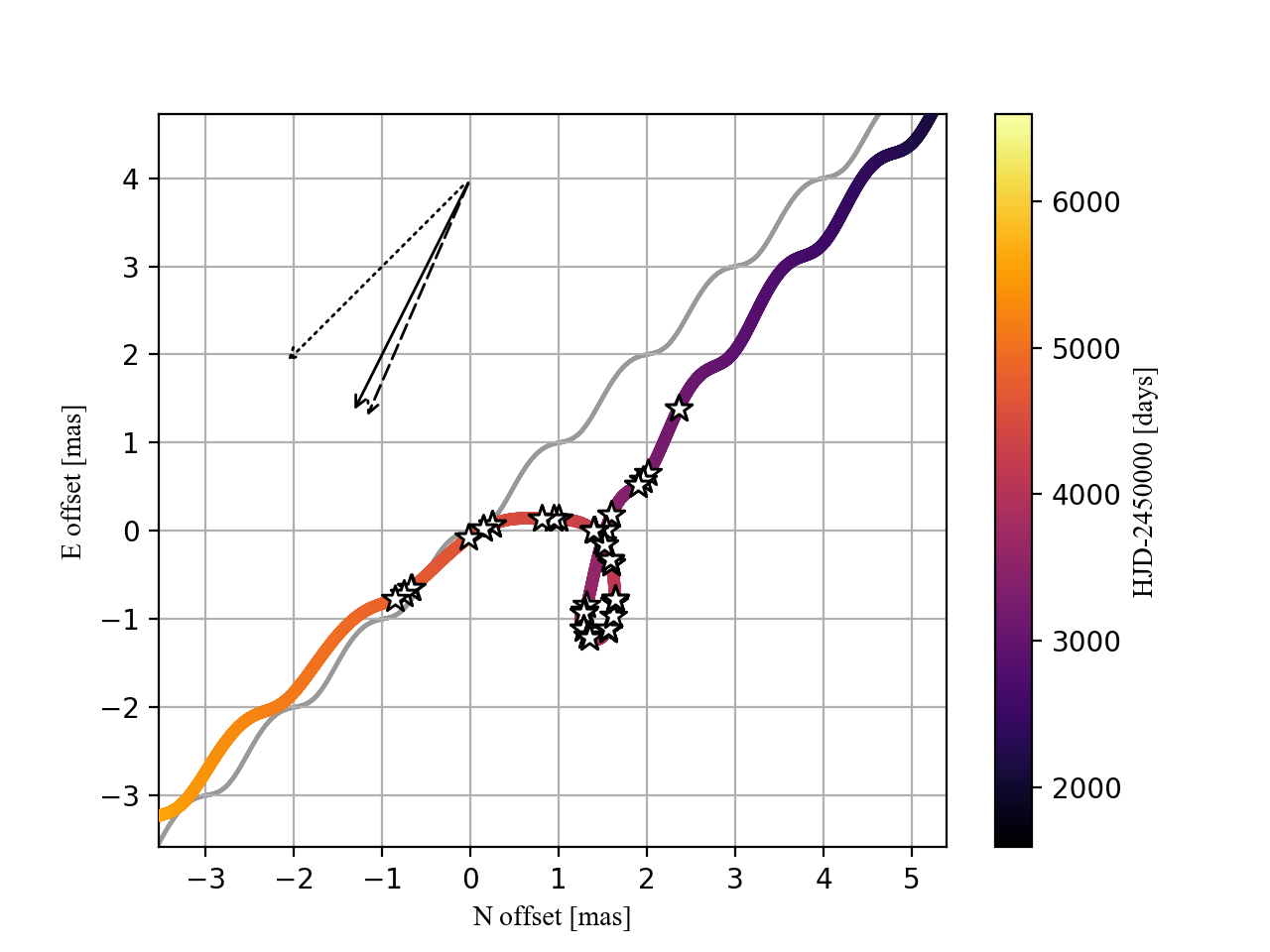}
\end{tabular}
\caption{Ideal motion curves with Gaia sampling for four different "classes" of curves for PAR-02 event with mass of the lens set to be a 8.6 \msun black hole. The four classes differ solely by the directions of the components of the proper motions (lens and source). Thus the lens mass, Einstein ring angular size, relative proper motion and other crucial parameters are the same for each panel. Dotted, solid and dashed arrows represent directions of source, lens and relative proper motions respectively. Length of the arrows is normalized so they only contain information about the direction, not the value of proper motions.
}

\label{fig:fourclasses}
\end{center}
\end{figure*}

To produce the astrometric trajectory of the centroid, the proper motions of the source and the lens are requisite. 
We pick and fix the lens-source relative proper motion at value $\mu_{rel}\approx8~mas/yr$ so that the mass of the lens is definitely BH-like and is consistent with the estimations from \cite{2016MNRAS.458.3012W}. 
Note, that for a particular value of $\mu_{rel}$ there is a large number of possible combinations of components of $\vec{\mu}_{src}$ and $\vec{\mu}_{lens}$ vectors that results in the same value of relative proper motion vector and thus the same mass of the lens.
Examples of different types of trajectories with over-plotted Gaia sampling points are shown on Figure \ref{fig:fourclasses} and the full list of parameters used to generate them is presented in Table \ref{tab:four_traj}.
\begin{table*}
\begin{center}
\begin{tabular}{|c|c|c|c|c|c|c|c|c|c|}
\hline
trajectory No.&$\mu_{src_N}$&$\mu_{src_E}$&$\mu_{len_N}$&$\mu_{len_E}$&$\mu_{rel}$&$M$&$\theta_E$&$D_l$&$D_s$ \\
$[-]$&$[mas/yr]$&$[mas/yr]$&$[mas/yr]$&$[mas/yr]$&$[mas/yr]$&$[M_{\odot}]$&$[mas]$&$[kpc]$&$[kpc]$\\
\hline
1. & $1.0$ & $1.0$ & $-2.22$ & $-6.42$ & $8.1$ & $8.6$ & $5.6$ & $1.7$ & $8.0$ \\
\hline
2. & $-1.0$ & $1.0$ & $-4.22$ & $-6.42$ & $8.1$ & $8.6$ & $5.6$ & $1.7$ & $8.0$ \\
\hline
3. & $1.0$ & $-1.0$ & $-2.22$ & $-8.42$ & $8.1$ & $8.6$ & $5.6$ & $1.7$ & $8.0$ \\
\hline
4. & $-1.0$ & $-1.0$ & $-4.22$ & $-8.42$ & $8.1$ & $8.6$ & $5.6$ & $1.7$ & $8.0$ \\
\hline
\end{tabular}
\end{center}
\caption{Sets of parameters for the four trajectories presented on the Figure \ref{fig:fourclasses}. Trajectories can differ significantly just due to the configuration of the particular proper motions of the source and the lens. In our analysis we did simulations for all trajectories and we haven't found significant difference in the results. Thus, in the following part of the work we only present one of them (number 2.)}
\label{tab:four_traj}
\end{table*}
Note, that the time zero point of mock Gaia observations is shifted to match the time of the event so its peak at $t_0$ occurred in the middle of such mock Gaia mission. 

\begin{figure*}
\begin{center}
\begin{tabular}{cc}
\includegraphics[width=\columnwidth]{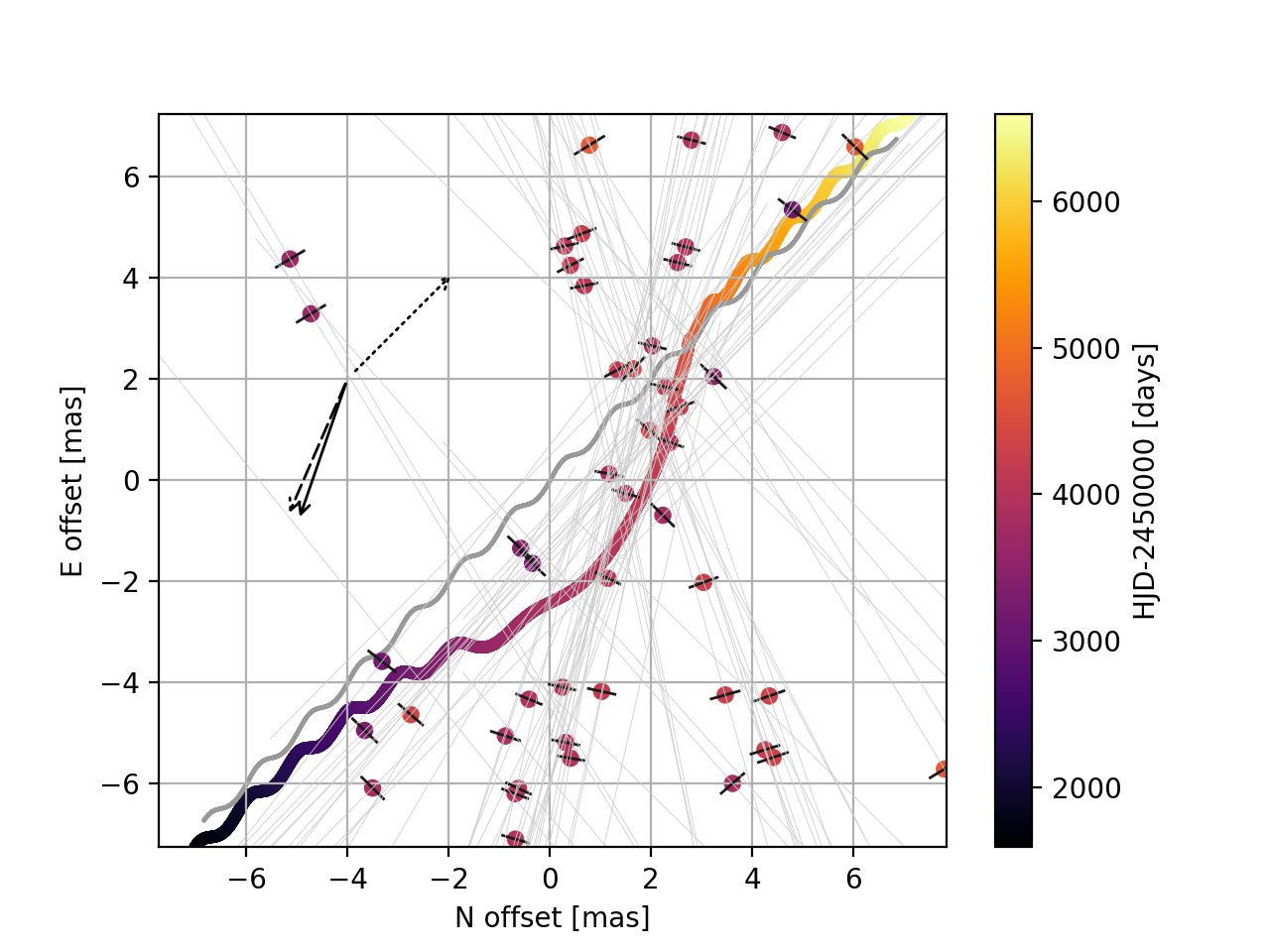}&
\includegraphics[width=\columnwidth]{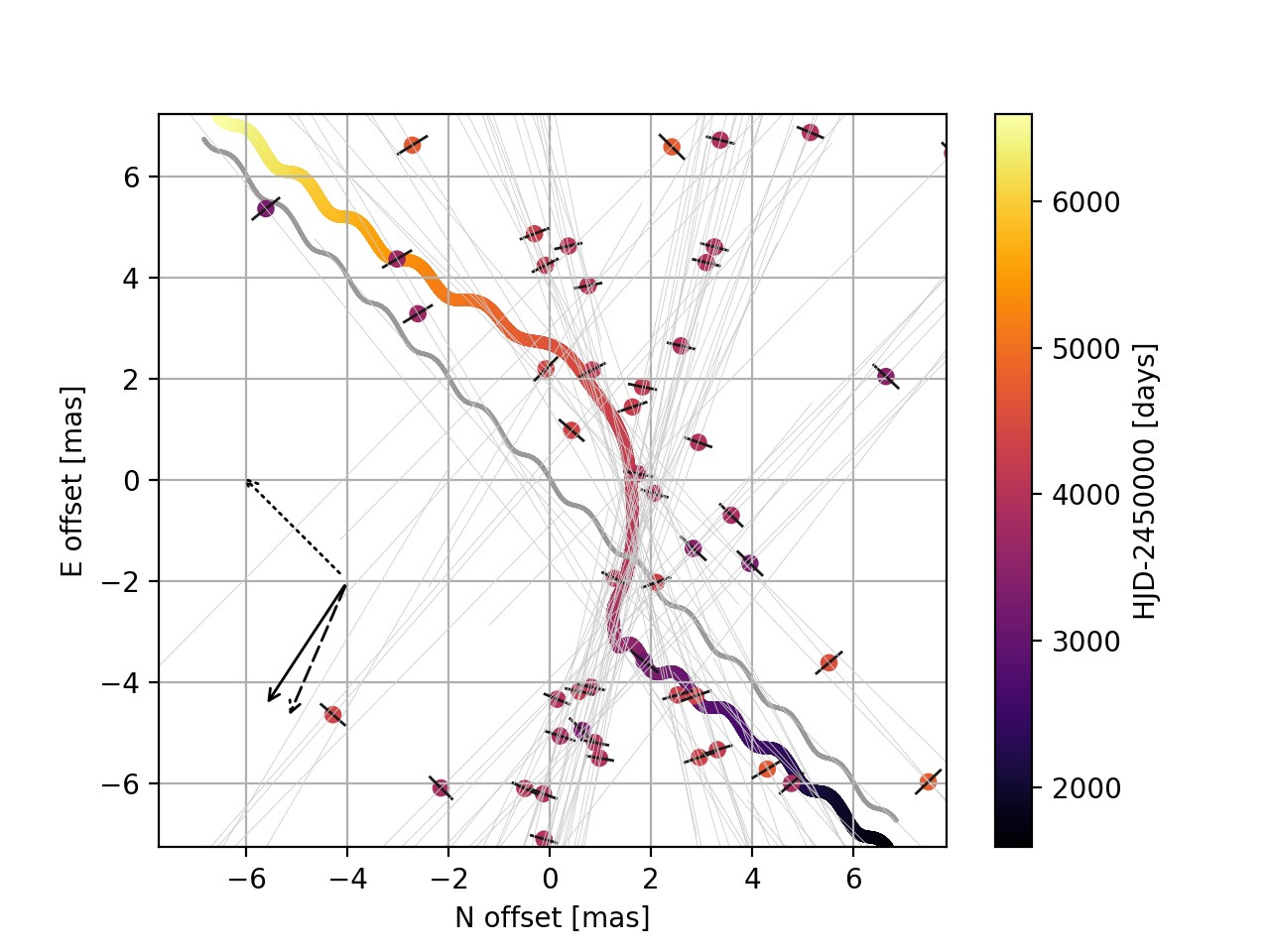}\\
\includegraphics[width=\columnwidth]{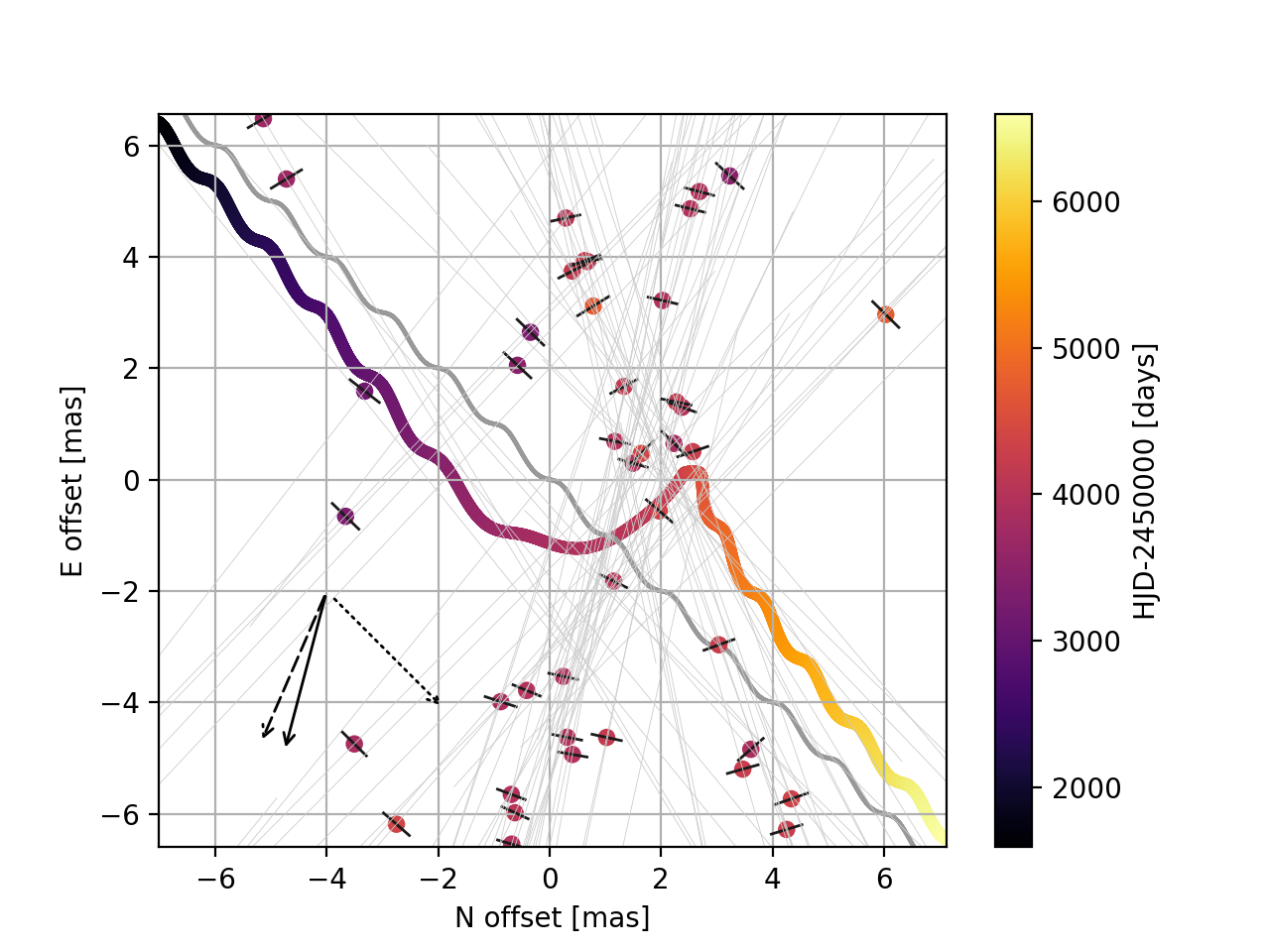}&
\includegraphics[width=\columnwidth]{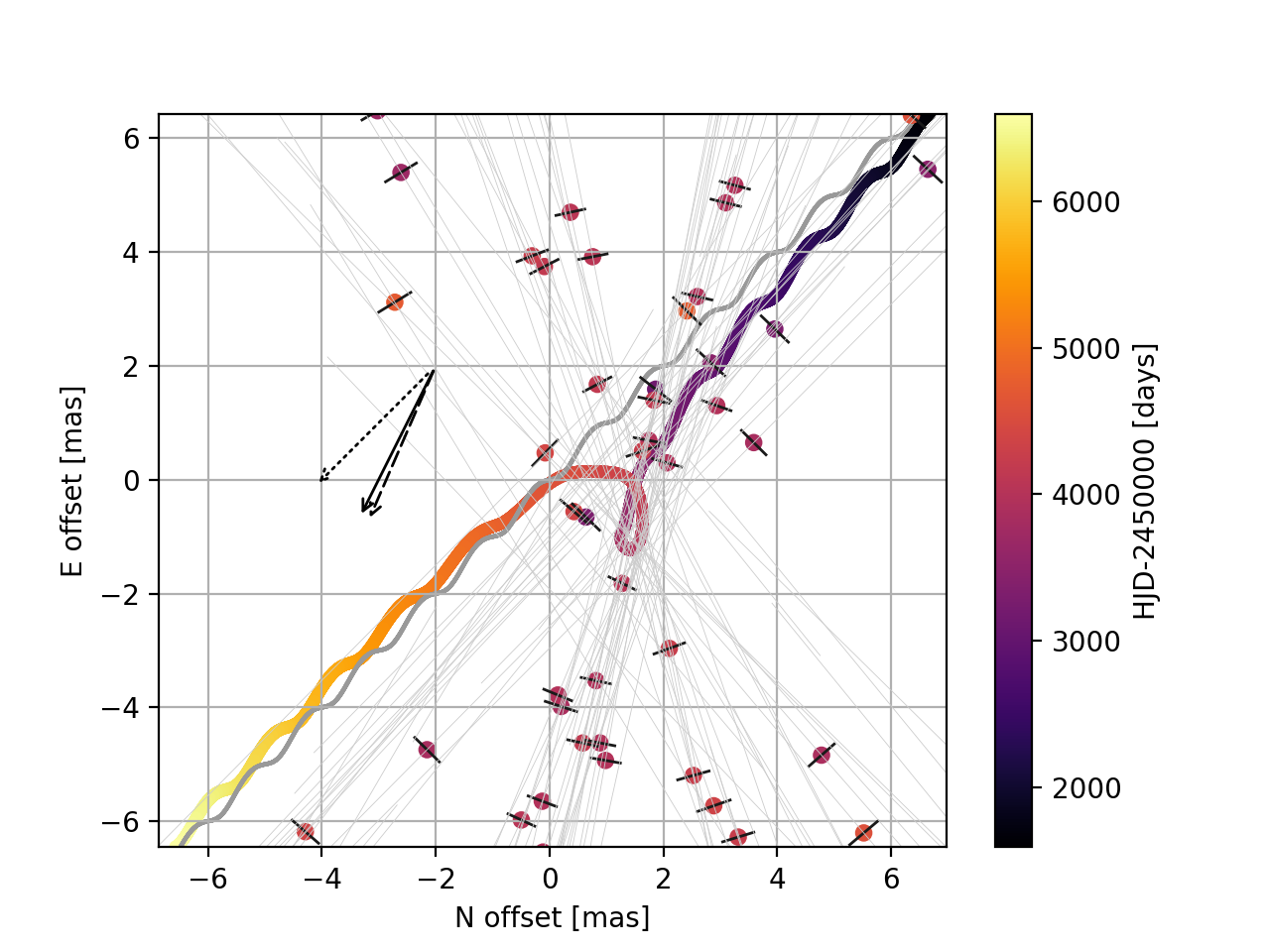}
\end{tabular}
\caption{Mock Gaia astrometric data points with realistic error-bars and scatter for four different "classes" of motion curves for black hole candidate PAR-02, parameters as in Fig. \ref{fig:fourclasses}.
The thick solid line shows the best fit trajectory obtained in our MCMC model when combining astrometry and photometry. Dotted, solid and dashed arrows represent directions of source, lens and relative proper motions respectively. Length of the arrows is normalized so they only contain information about the direction, not the value of proper motions.
}
\label{fig:simul}
\end{center}
\end{figure*}

\begin{figure*}
\begin{center}
\includegraphics[width=\textwidth]{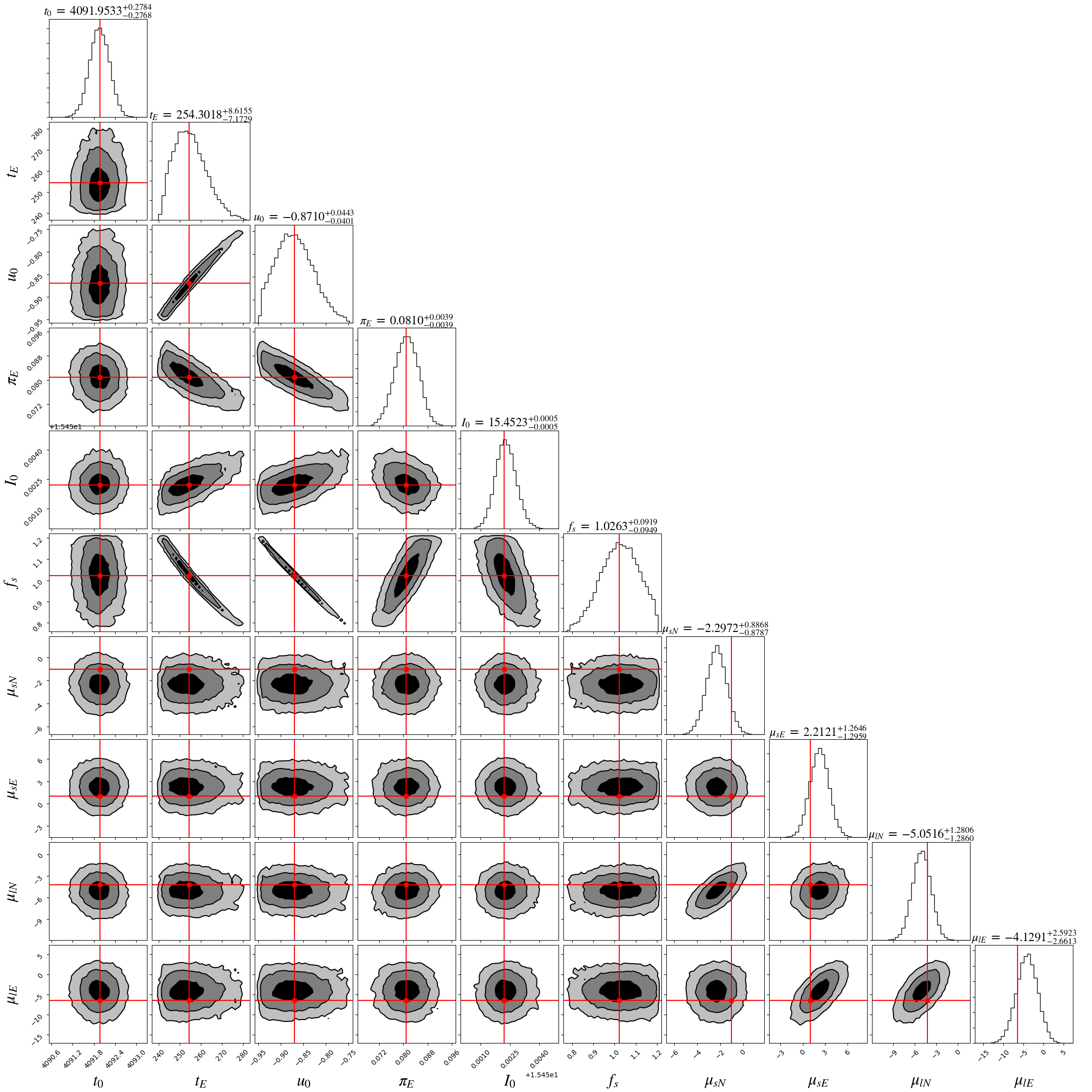}
\caption{Parameters distributions from the MCMC simulations for combined OGLE photometry and mock Gaia astrometry for black hole candidate lens PAR-02 for assumed mass of about 8.6 \msun. Shown here is the first of the two solutions found. The red vertical lines mark the real parameters values - ones that were used to generate mock Gaia astrometry (see Table \ref{tab:model} and \ref{tab:four_traj}). The distributions for the first seven parameters reproduce  these values very well, while for the proper motions they are slightly off. The reason is that information about $t_E, u_0, t_0, f_s, I_0$ and microlensing parallax can be derived from the lightcurve, very well sampled with the OGLE data. On the other hand, the proper motions are constrained mostly by the mock Gaia astrometric data, rather poor for the PAR-02 brightness ($I=15.5$ in baseline).
}
\label{fig:mcmc}
\end{center}
\end{figure*}

We then simulated the mock Gaia astrometric data for four different combinations of vectors $\vec{\mu}_{src}$ and $\vec{\mu}_{lens}$ adding the expected noise to the positional measurements, as described in Section \ref{sec:gaia}.
Figure \ref{fig:simul} shows Gaia data points expected for PAR-02 microlensing event. 
Note, the error-bars are significantly smaller in the AL direction than AC. 
At this point, after simulating astrometric data for Gaia, we began the procedure we would run in the future, with both real OGLE and Gaia data available, and used the combined photometric and astrometric model for light and motion curve data.
As the geometry of the trajectory significantly varies depending on the proper motions components we calculated the solution for four different sets of proper motions. 
Each of these configurations yield the same relative proper motion and mass of the lens as $\mu_{rel}=8.1~mas/yr$ and $M=8.6~M_{\odot}$ but has noticeably different geometry. 
Nevertheless, the resulting mass distribution and thus the method accuracy does not vary significantly from one case to another. 
The example and typical MCMC results of the modelling are shown in Fig. \ref{fig:mcmc}, with the red lines marking the values of the parameters input in the simulation. 

\begin{figure*}
\begin{center}
\includegraphics[width=\textwidth]{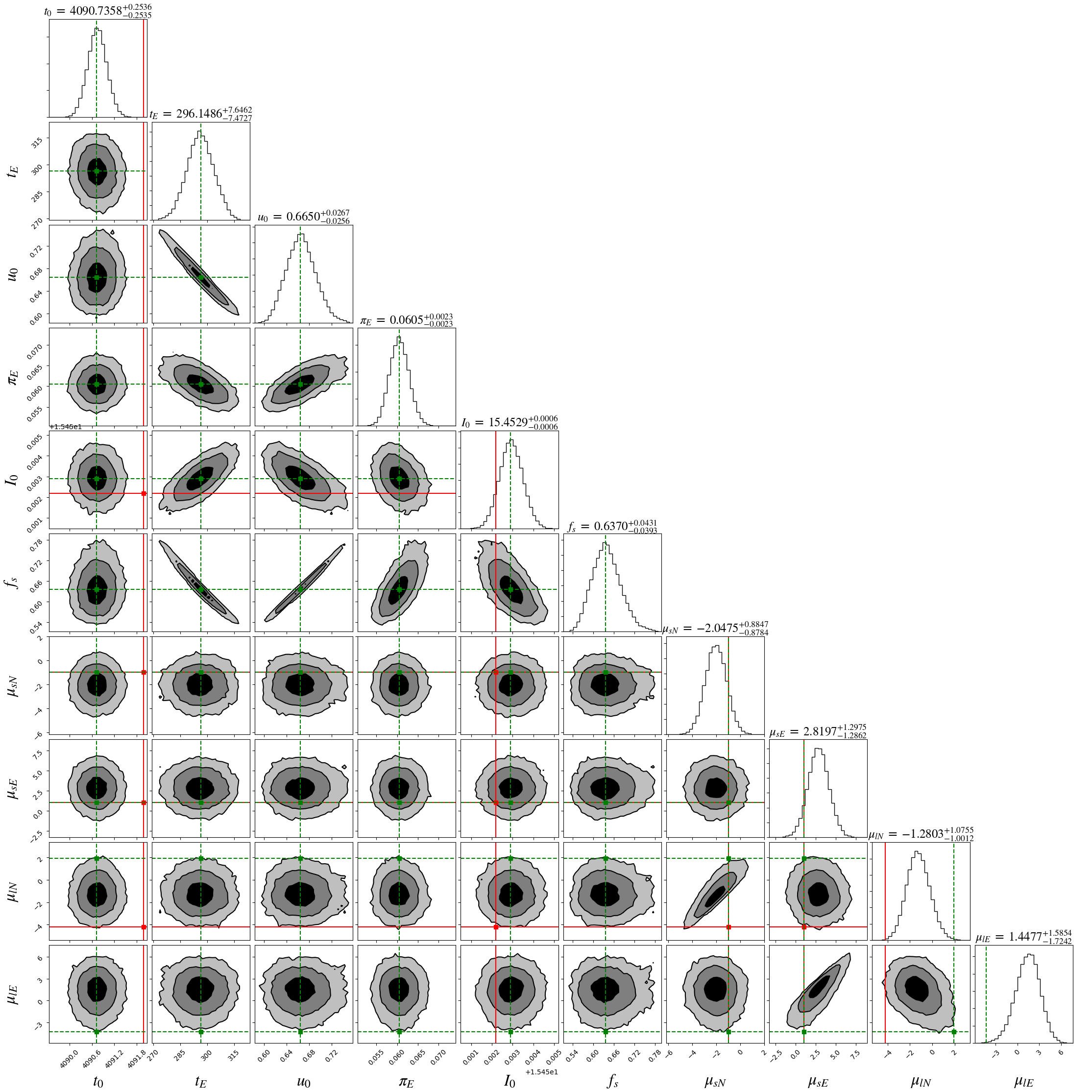}
\caption{As in Fig. \ref{fig:mcmc}, for the second solution found. We know that this solution is not the correct one, as parameters do not agree with the ones used in simulations of the data. Nonetheless, in the realistic situation it wouldn't be possible to unambiguously reject this solution. The green dashed vertical lines mark the parameters values for the second photometric solution. Note that for the photometric part (first seven parameters), this second result of the joint MCMC simulations is reproducing the values of parameters from the second photometric solution (see Table \ref{tab:model}).}
\label{fig:mcmc2}
\end{center}
\end{figure*}

Joint MCMC modelling (for both astrometric and photometric data) has also shown, that there is a second likely solution. 
This solution's MCMC distributions are shown in Fig. \ref{fig:mcmc2}. Existence of the second solution is not surprising, as this minimum was already found in the photometry modelling (Table \ref{tab:model}). 
Indeed, for this combined data solution all resulting distributions for seven photometric parameters agreed with the second photometric solution. On the other hand, the proper motions distributions, especially for the lens, although converged, did not reproduce the true input values. In consequence, the related physical parameters, primarily the mass of the lens and its distance, did not agree with the the input values.

We note that for the brightness of PAR-02 event at a level of $I=$15.5 mag at the baseline, the degeneracy between solutions remains and is not broken by additional information from the astrometry. However, as we show later, for brighter magnitudes it can be broken due to the fact that we measure the relative proper motion accurately(\eg  \citealt{2014ApJ...784...64G}).

\begin{figure}
\begin{center}
\includegraphics[width=\columnwidth]{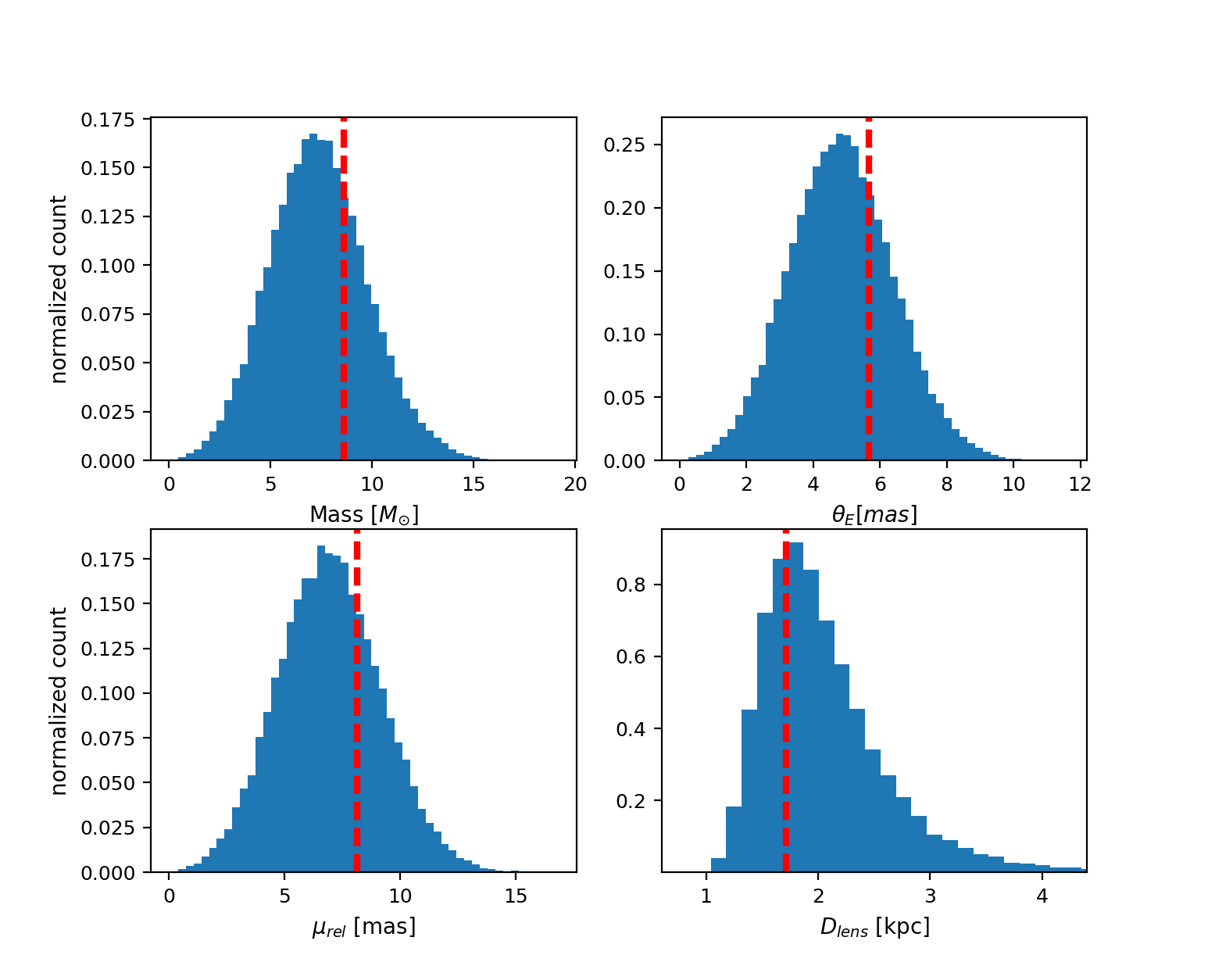}
\caption{Distributions of the secondary parameters, derived from the posterior distributions presented on Fig. \ref{fig:mcmc}. The red line shows the input value. The standard deviation of the mass measurement $\sigma_M \approx 2.3 M_{\odot}$, \ie about 30 per cent.}
\label{fig:mcmcmass}
\end{center}
\end{figure}

\begin{figure}
\begin{center}
\includegraphics[width=\columnwidth]{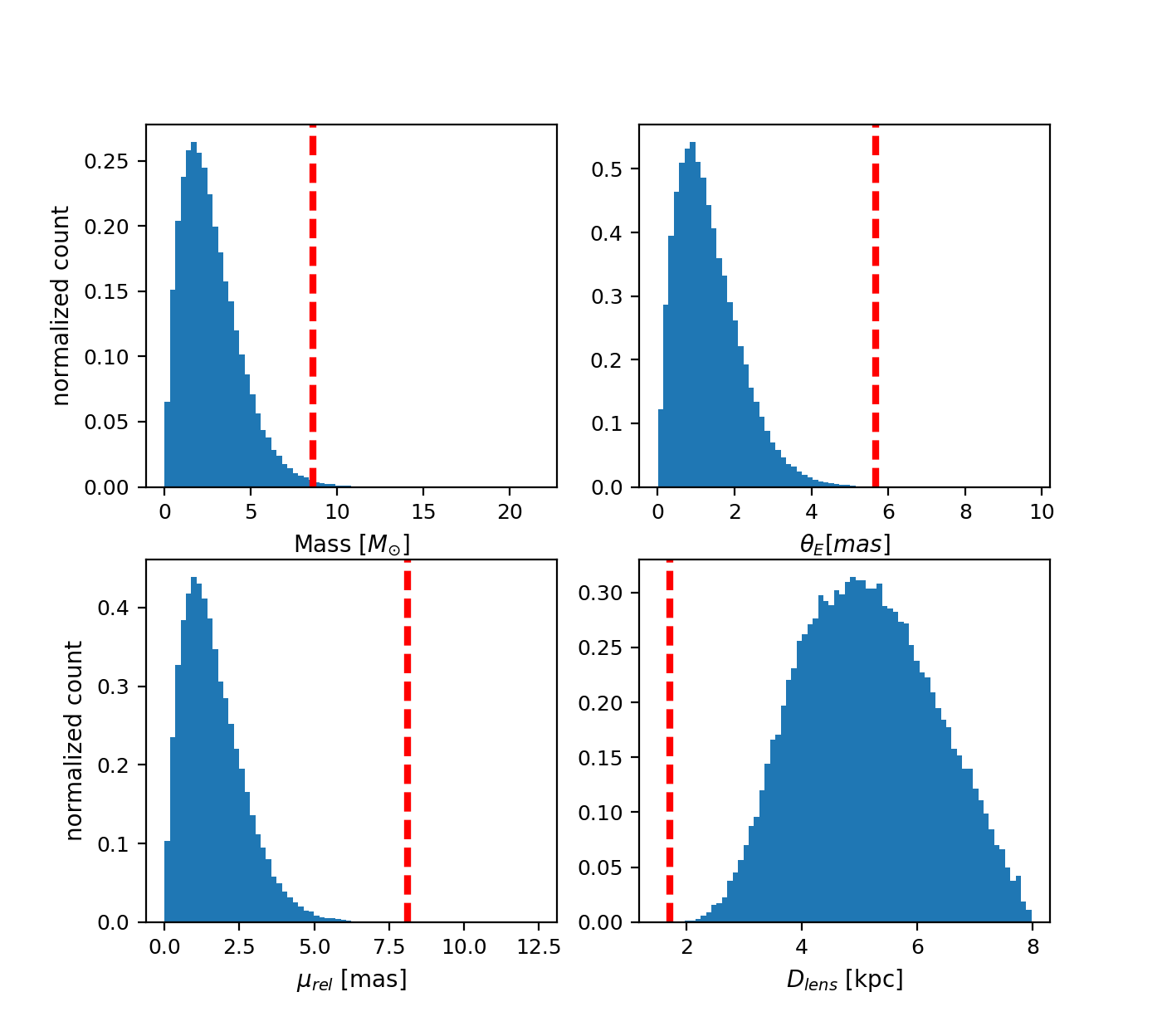}
\caption{Distributions of the secondary parameters, similar to the Fig. \ref{fig:mcmcmass}, but derived from posterior distributions for the second solution, presented on Fig. \ref{fig:mcmc2}. Even though the fit converged, the input parameters are not reproduced here. Thus, in the real-life situation, we wouldn't be able to unambiguously determine the mass of the lens, as degeneration known from the photometric data remains.}
\label{fig:mcmcmass2}
\end{center}
\end{figure}

For each of the solutions the mass and the distance can be computed based on the parameters of the microlensing fit. 
The resulting distributions are displayed in Fig. \ref{fig:mcmcmass} and Fig. \ref{fig:mcmcmass2}, along with the retrieved proper motion and Einstein Radius size.

In the first solution the mass of the lens in PAR-02 for the proper motions assumed for its most likely value was retrieved with a standard deviation of about 2.3 $M_{\odot}$, \ie about 30 per cent.
The distance to the lens was obtained with somewhat better accuracy, \ie $\sigma_{D_l}\approx$0.5 kpc for 1.95 kpc yielding about 25 per cent. As for the second solution, the mass distribution gives accuracy of mass determination worse than 50 per cent, while the relative error for the distance is again about 25 per cent.

\begin{figure*}
\begin{center}
\begin{tabular}{cc}
\includegraphics[width=\columnwidth]{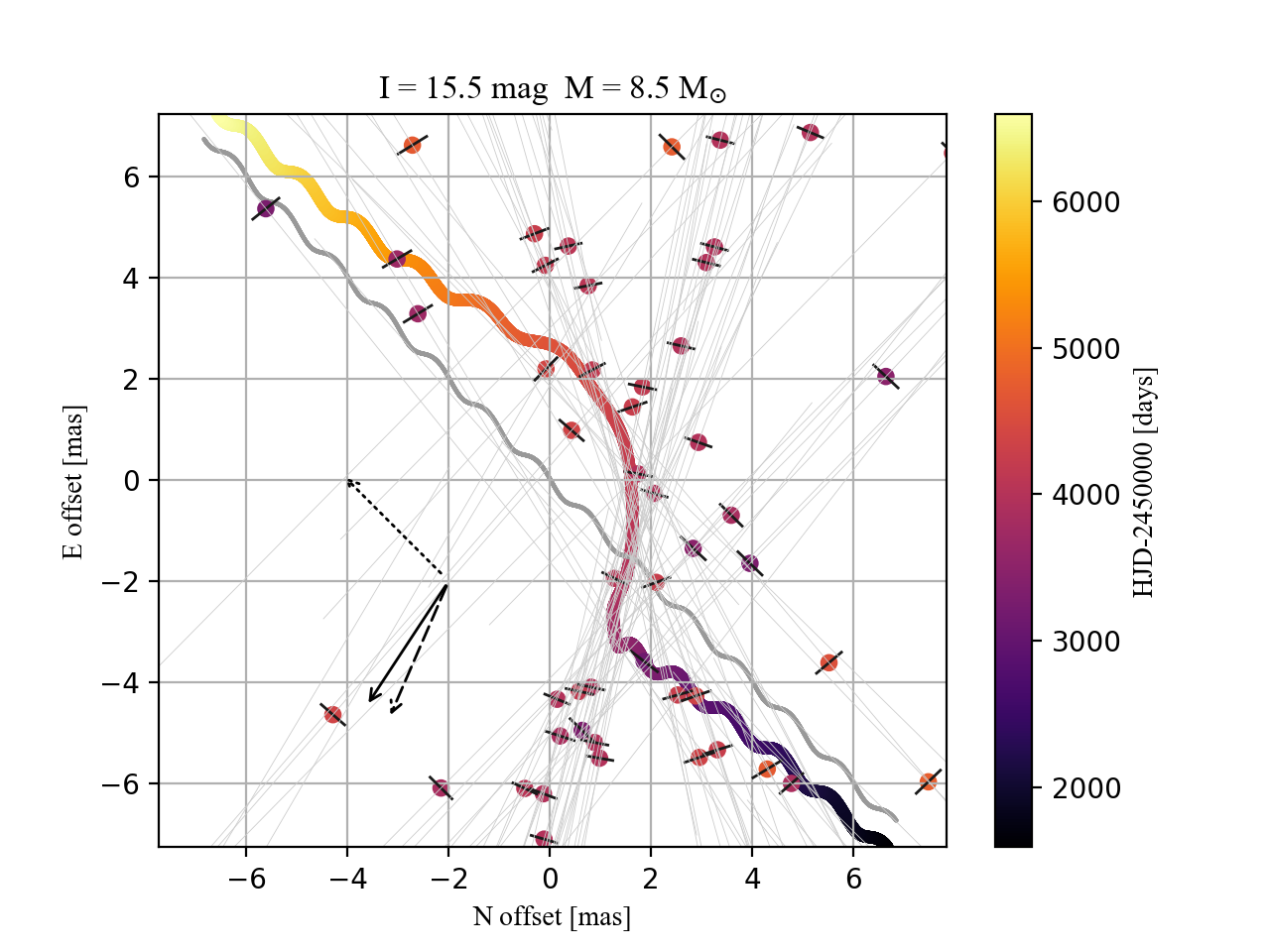}&
\includegraphics[width=\columnwidth]{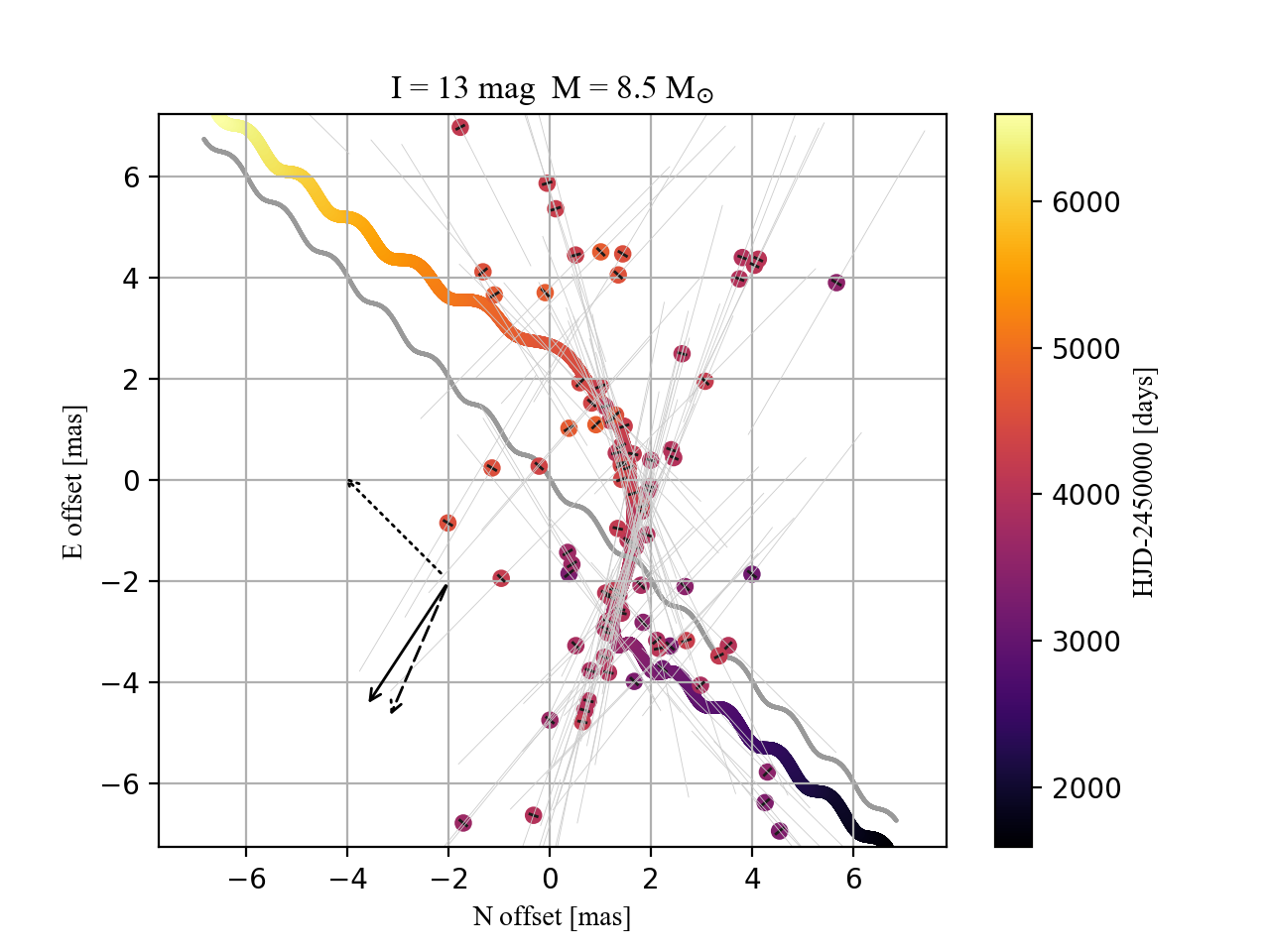}\\
\includegraphics[width=\columnwidth]{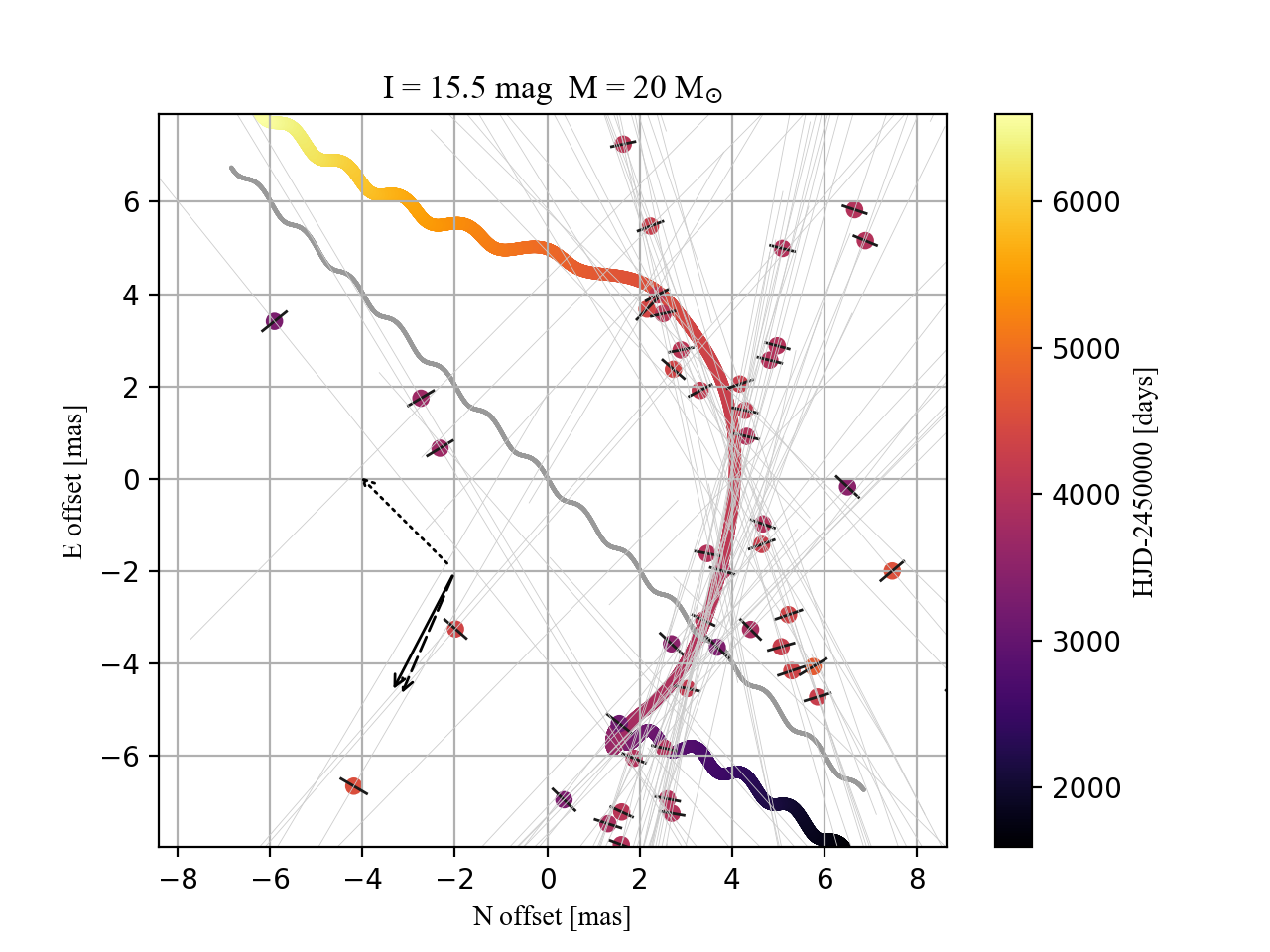}&
\includegraphics[width=\columnwidth]{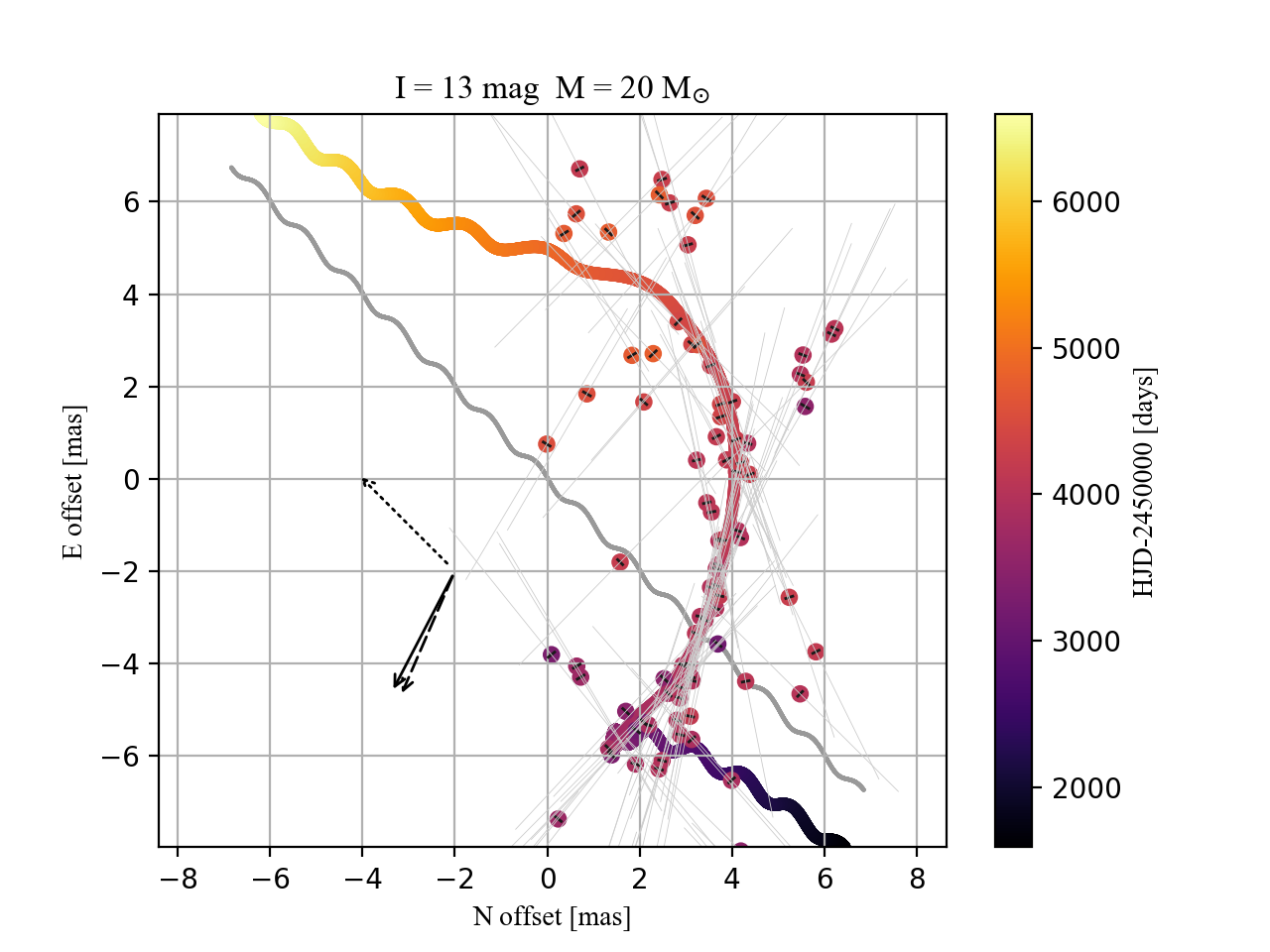}
\end{tabular}
\caption{
Mock Gaia astrometric data points with realistic error-bars and scatter. Panels differ with baseline magnitude ($I$=15.5 and 13 mag top to bottom, respectively, and mass: 8.6 and 20\msun for left and right, respectively. 
The notation and markings are as in Fig. \ref{fig:fourclasses}. 
}
\label{fig:simulmag-mass}
\end{center}
\end{figure*}

Next, we used our event PAR-02 to test how well the mass of the lens is going to be retrieved for more heavy lenses and for different brightness of the events. 
Figure \ref{fig:simulmag-mass} shows the simulated astrometric data of Gaia for different baseline magnitudes and masses of the lens.  

\begin{figure*}
\begin{center}
\includegraphics[width=15cm]{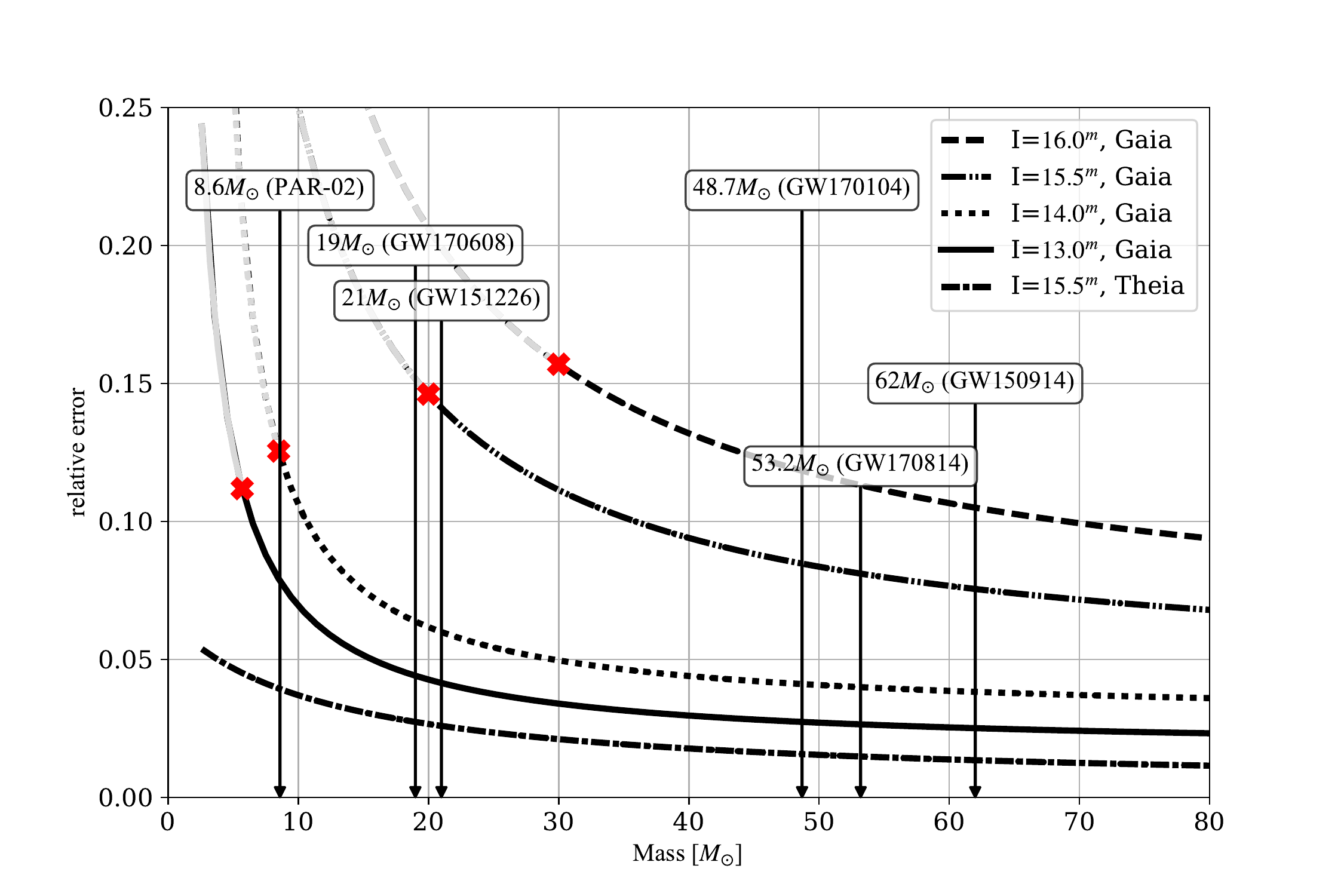}
\caption{Relative error $\frac{\Delta M}{M}$ in the lens mass determination for different masses of the lens and different observed magnitudes of the microlensing event. The vertical lines shows the position of the most likely parameters of our prototype BH lens, PAR-02 and masses of BHs recently measured in the gravitational waves detections. Different lines are shown for different baseline magnitudes. Red cross marks the transition point between the degenerated and non-degenerated areas. The black part of the curve is a region where
the degeneracy can be broken with the additional information from astrometry. The grey part of the curve represents the region, where the baseline is too faint and/or the lens too light for the astrometry to provide enough information to brake the degeneracy.
Dash-dotted line represents the accuracy that can be achieved by the next generation astrometric missions. We recall, that for the PAR-02 event $G-I\approx 1.3$ mag.
}
\label{fig:masserror}
\end{center}
\end{figure*}

In order to compute the mass of the lens, one needs both astrometry and photometry. 
For a fixed set of microlensing parameters for PAR-02, we generated a series of astrometric data sets, each for increasing $\theta_E$. We adopted the strategy to use the real photometry in our simulations, which means we don't want to change the lightcurve. Thus, to change the $\theta_E$ and mass of the lens we changed the length of the relative proper motion vector, $\mu_{rel}$ (see equation 1.).
We combined the simulated astrometric data sets for different \thetaE with the original OGLE photometry of PAR-02. 
Since we modified the relative proper motion, the time-scale $t_E$ of the event was unchanged and the photometry was used in its original form. 
For each mass (proper motion) we generated mock astrometric Gaia data and then run the MCMC model on combined astrometry and photometry data to find all parameters and their errors for such simulated microlensing event. 
Note that in this exercise we used so-called "uninformative" priors, which are simply flat distributions for all parameters.  
Figure \ref{fig:masserror} shows the $\sigma$ on mass computation for a black hole lens in a microlensing event, similar to PAR-02 in terms of photometry (brightness and duration), with increasing mass. Only the mass derived for the solution closest to the input values is shown, since $\sigma$ on mass remains similar for both solutions. 
However, we marked the point where the degeneracy between the two solutions can be broken with added astrometry (red crosses).
The location of the crosses is dictated by the statistical significance of the second (wrong) solution. We reject the hypothesis that the second solution should be taken into account and the degeneracy remains, for all models resulting in $\chi^2$ such as the $p-value$ is higher than our significance level $\alpha=0.003$. For this test we only use fit (and $\chi^2$ value) of the motion curve, as deviation of the photometric model is insignificant compared to the astrometric one.

We also tested how the mass accuracy measurement changes for events brighter and fainter than PAR-02.
Again, for each mass (proper motion) we generated a set of astrometric mock Gaia data, but the observed magnitude was modified, therefore the scatter and errors in Gaia data were adjusted as described in Section \ref{sec:gaia}. 
OGLE-III photometry was also shifted and its scatter and measurement errors were adjusted following the procedure described in \cite{2011MNRAS.416.2949W}. 
In Fig. \ref{fig:masserror} different lines show the accuracy of lens mass measurement for different baseline magnitudes.

Additionally, we repeated the test on the accuracy of lens mass determination for the proposed next generation astrometric mission\citep{2016SPIE.9904E..2FM}. We assumed that such mission had a significantly improved accuracy of its astrometry compared to Gaia, reaching 100 $\mu$as at 15 mag and the sampling similar to that of Gaia. 

\section{Event rates}
In this section we discuss the theoretical rate of astrometric microlensing events caused by 
stellar-origin BHs that could potentially be observed towards the Galactic bulge. 
Our intention is to obtain a rough estimate of such rate, disregarding factors that would vary the result 
by less then a factor of $\sim 5$. A more accurate calculation, together with a study of the Galactic population 
of stellar-origin BHs, will be presented in a separate paper. 

Following \citet{Dominik2000}, the probability of an astrometric microlensing event 
occuring for a given source located at distance $D_S$ during an observation time 
$T_{\rm obs}$ and causing a centroid shift greater than $\delta_T$
is given by their equation (68), here written before integrating over $M$:
\begin{equation}
\label{eq:gamma}
\begin{aligned}
 \gamma = {} & 4 \sqrt{\frac{G}{c^2}} D_S T_{\rm obs}^{3/2} v^{3/2} \delta_T^{-1/2} \\
 & \quad \quad \times \int_{M = 0}^{\infty} \int_{x=x_{\rm min}}^{x_{\rm max}}  \eta(x) \sqrt{M} \sqrt{1-x} ~ f(M) dx dM ~~~ ,
 \end{aligned}
\end{equation}
where $v$ is the relative source-lens velocity, 
$x = D_L / D_S$ where $D_L$ is distance to the lens, $M$ is the mass of the lens, $\eta(x)$ is the number density of considered lenses
(here stellar BHs in the disc)\footnote{Note that \citet{Dominik2000} use
mass density of considered lenses $\rho(x)$ divided by the lens mass $M$ instead.}
and $f(M)$ is their mass function (normalized to unity). Limits in the integral over $x$ are such that for a lens with a given mass $M$ the centroid shift is larger than $\delta_T$.

In order to estimate the number density of BHs 
in the Milky Way disc $\eta_{\rm BH}$, we utilize the \textsc{StarTrack} population 
synthesis code \citep{Belczynski2002, Belczynski2008} to generate 
a population of stellar-origin BHs. 
We evolve $\sim 7.4 \times 10^6$ binary systems, each initially comprised 
of two zero-age main sequence stars of solar metallicity \citep[$Z_{\odot} = 0.02$,][]{Villante2014}, 
with the primary component being a massive star of $M_a \geq 20 M_{\odot}$ 
and thus a likely BH progenitor \citep{Fryer2012}.
We draw the initial binary parameters 
from the distributions of \citet{Sana2012}, we utilize the Kroupa-like initial mass function 
\citep{Kroupa1993} with a power-law exponent for the massive stars $\alpha_3 = 2.3$ \citep{Bastian2010}, 
and we assume 100\% binary fraction for massive primaries $M_a \geq 10 M_{\odot}$ and 50\% for 
$M_a < 10 M_{\odot}$ \citep[multiplicity of stars increases with their mass, see][]{Duchene2013}.
We normalize our simulated population to a 
constant star formation rate for the Milky Way disc of $3.5 M_{\odot} yr^{-1}$ over the last 10 Gyr.
We emphasize that rates discussion in this paragraph does not refer to the massive BHs similar to those detected in gravitational waves experiments, as they were most likely formed in very low metallicity stellar populations \citep[$Z < 0.1 \, Z_{\odot}$, eg.][]{2016Natur.534..512B}, which are not included in our simulations (most of the star formation in the Milky Way disc happened in $Z \sim Z_{\odot}$, \citealt{Robin2003}).

All the BHs in our simulations can be divided into three categories: (a) BHs that survive as components 
of binary systems, (b) single BHs originating from binaries that were disrupted during supernovae explosions, and (c) 
single BHs originating from single stars that were formed in mergers of binary components.
With regards of the last group, 
we assume that a merger of two stars with masses $M_1$ and $M_2$ such that $M_1 \geq M_2$ will result in 
a formation of a BH if $M_1 + 0.5 M_2 \geq 21.5 M_{\odot}$, where $\sim 21.5 M_{\odot}$ 
is the minimal initial mass that a star needs to have 
in order to become a BH for solar metallicity in \textsc{StarTrack}. 

The numbers of BHs in each of the three categories are: (a) $0.56 \times 10^6$, (b) $4.91 \times 10^6$ and (c)
$2.19 \times 10^6$, respectively. Notice that BHs that still reside in binary systems are a minority
($\sim 7\%$ of all the BHs),
which is why, for the purpose of this estimate, we assume that all the lenses are single objects. 
We wish to highlight that the number of binaries disrupted in supernovae explosions is subject
to the assumption on the BH natal kicks. Here, we utilize the ''Rapid'' supernova model of \citet{Fryer2012} 
with the magnitude of natal kicks reduced due to fallback. As a result, about $\sim 56\%$ of the BHs in sollar metallicity 
are formed in direct collapse (i.e. no natal kick).

The overall number of BHs (both single and in binaries) in our simulation is $\sim 7.7 \times 10^6$. When normalized, 
this corresponds to $N_{\rm tot} = 6.24 \times 10^8$ potential BH lenses formed in the Milky Way disc, and majority of them (93\%) are single objects. We assume that their spatial distribution 
follows the mass distribution of the thin disc in the synthetic model of \citet{Robin2003}:
\begin{equation}
\eta_{\rm BH} (R,z) = \eta_0 ~\Bigg({\rm exp}\Bigg(-\sqrt{0.25 + \frac{a^2}{h_{\rm R+}^2}}\Bigg) - {\rm exp}\Bigg(-\sqrt{0.25 + \frac{a^2}{h_{\rm R-}^2}}\Bigg)\Bigg) ,
\end{equation}
where $a^2 = R^2 + z^2 / \epsilon^2$, $\eta_0$ is a normalization constant, 
and the values of parameters are $\epsilon = 0.079$, $h_{\rm R+} = 2530 \, \rm pc$ and
$h_{\rm R-} = 1320 \, \rm pc$.

Most of the BHs in our simulations have masses between $5.5$ and $9 \, M_{\odot}$, with the tail of the distribution reaching up
to $\sim 15 \, M_{\odot}$, and the mean mass of about $7.5 \, M_{\odot}$. This is in agreement with the dynamically inferred 
masses of BHs in Galactic X-ray binaries \citep{Tetarenko2016}.

Having all the ingredients, we are able to evaluate Eq.~\ref{eq:gamma} for $\gamma$ in the direction of observation towards 
the Galactic bulge. The event rate 
$\Gamma$ is given by $\Gamma = \gamma ~ T_{\rm obs}^{-1}$:
\begin{equation}
\begin{aligned}
 \Gamma = 4.65 \times 10^{-7} (1 - f_{\rm esc}) \Bigg( \frac{D_S}{8 \, \rm kpc} \Bigg) \Bigg( \frac{T_{\rm obs}}{5 \, \rm yr} \Bigg)^{3/2} \\
 \times \Bigg( \frac{v}{100 \, \rm km \, s^{-1}} \Bigg)^{3/2} 
 \Bigg( \frac{\delta_T}{1 \, \rm mas} \Bigg)^{-1/2}  \rm yr^{-1},
 \end{aligned}
\end{equation}
where $f_{\rm esc} \lesssim 0.6$ is a fraction of BHs from disrupted systems which potentially left the bulge due to high natal kicks. For the bulge sources ($D_S \approx 8 \, \rm kpc$), typical relative velocities of $\sim 100 \, \rm km \, s^{-1}$,  $f_{\rm esc} = 0$,
five years of observation time and a realistic threshold centroid shift of $1 \, \rm mas$ required for detection, the 
event rate per observed source is about $4 \times 10^{-7} \, \rm yr^{-1}$. Notice, that the order of magnitude is in 
agreement with the results of \citet{Dominik2000}, who would obtain $2.7 \times 10^{-6} \, \rm yr^{-1}$ for $\delta_T = 1 \, \rm mas$ (see their Table 3)
using mass density that is an order of magnitude larger than ours for only BHs.

If we assume that about $10\%$ of the sources reported in the Gaia Data Release 1 are resolved and located in the Galactic 
bulge, we could expect of the order of $\sim 10^8$ potential sources \citep{GaiaDR1}. About 5\% ($\approx 5\times 10^6$) of them are brighter than $G=15.5$mag. For this subset, Gaia astrometric time-series precision will be relatively good and the degeneracy between multiple microlensing solutions broken (see figure \ref{fig:masserror}). Assuming that the event rate is $4 \times 10^{-7} \, \rm yr^{-1}$ and $5\times 10^6$ bright sources will be observed for the duration of five years, we estimate that the order of few astrometric microlensing events caused by stellar-origin BHs could be observed by Gaia.


\section{Discussion}
In this work we have considered a single microlensing event, OGLE3-ULENS-PAR-02, as a best example and a prototype for a black hole lens candidate.
Based on the real photometry from OGLE-III from years 2001-2008 and known characteristics of the Gaia mission, we simulated the astrometric motion curve as it will be measured by Gaia for similar events. 
Simultaneous fitting of the photometric and astrometric models to both sets of data resulted in the probability distributions for the event parameters, including mass and distance of the lens. 
As shown in \cite{2016MNRAS.458.3012W}, the most likely mass of the lens in PAR-02 event (hence the black hole mass) is about $10~M_{\odot}$, we simulated Gaia astrometry that reproduces such mass. 
We found that the mass can be retrieved from the combined data set with accuracy of about 30 per cent, however, for the brightness of PAR-02 at the baseline (15.5 mag in OGLE $I$-band, equivalent to about 16.8 mag in Gaia $G$-band), the degeneracy between the two solutions remains. 
It means that despite we will be able to derive the mass of the lens fairly accurately, in order to recognise which of the solutions is the physical one, we still require additional observations to break the degeneracy (\eg high resolution imaging or estimation of the source proper motion).


We then estimated how well other masses can be measured. 
By modifying the relative proper motion parameter we changed $\theta_E$, \ie the size of the astrometric microlensing ellipse, therefore we tested the behaviour of the astrometry with changed mass of the lens. 
The larger the Einstein Radius, the more significant astrometric signal, hence naturally, the accuracy of measurement of the mass improved with mass of the lens (Fig. \ref{fig:masserror}). 
For small masses, between 5 and 10 $M_{\odot}$, the error in mass measurement for the brightest events (I$<$13 mag) was about 10 per cent and the degeneracy between solutions was clearly broken. For fainter baselines the problem with degeneracy remains and the mass can be measured with an error-bar of about 50 per cent. 


%

Even though the dynamically inferred masses of BHs in the Galactic BH X-ray binaries are generally around 10 $M_{\odot}$ and no larger than 20 $M_{\odot}$ \citep{Tetarenko2016}, there is no such measurements for systems in the Galactic disc (b ~ 0 deg) where observations of BH X-ray binaries are made difficult by extinction. Meanwhile, the most massive BHs are believed to be formed in direct collapse events with no natal kicks (e.g. \citealt{Fryer2012}), and, if that is the case, would most likely still reside in the star forming regions of the disc.

However, the most robust mass measurements will be possible for Galactic black holes with masses similar to these recently found to form in black hole mergers discovered in gravitational waves\citep{2016PhRvL.116f1102A}. 
That, in turn, should allow to estimate the merger rate in the Milky Way and compare with the observed signal \citep{2016Natur.534..512B}.
Even for the faintest baselines considered here ($I$-band $<$16 mag, equivalent of Gaia's $G$-band $<$17.3 mag), the degeneracy can be broken for masses above about 35 \msun and the mass measurement can be achieved with accuracy of around 10 per cent. 
Note, that in practice, the same value of $\theta_E$ (and mass) can be observed for a different sets of parameters $\mu_{rel}$ and $t_E$. 
The value of the angular Einstein Radius defines the amplitude in the astrometric path, however, proper motion and time-scale will influence the duration of the path, \ie how long it will take for the microlensing astrometric loop of the centroid to return to its original position. For very long lasting events the duration of the Gaia mission of 5 years might not be enough to cover a significant part of the astrometric loop and hence to reliably measure the size of $\theta_E$. 
Moreover, an increased $t_E$ means the event will last longer also in photometry. As shown in \cite{2015ApJS..216...12W}, the detection efficiency for very long events in 8-years long data of OGLE-III, starts to drop at time-scales longer than 1 year. This is due to the fact, that some of such events will evolve very slowly and might not be covered by the data span and might resemble slow variable stars, therefore their retrieval in the data is less efficient. Very long photometric time-series are required in order to detect long and massive events, an opportunity which is given by combined OGLE-III and OGLE-IV data (from 2001 for some parts of the sky). 

On the other hand, proper motions of lensing black holes or neutron stars can be higher than for typical stars due to natal kicks (\citealt{2013MNRAS.434.1355J}, \citealt{2012MNRAS.425.2799R}), which could be of the order of 100-200 km/s, corresponding to additional proper motion of 20-40 mas/yr for a lens at distance of 1 kpc.
It means, that the time-scales is such events can be similar or even shorter than in our prototype PAR-02. 
In such cases the astrometric and photometric signatures of an event will last shorter and for time-scales shorter than about 50 days it might be too short to robustly measure the microlensing parallax (unless it is measured with space observations, with, \eg Spitzer). However, even for relative proper motions of order of 30 mas/yr and $t_E\sim$ 50 days, the size of the Einstein Radius is still of order of 4 mas, therefore such amplitude should be measured with Gaia astrometry, at least for the brightest events. 

In 2002 Belokurov and Evans (\citealt{2002MNRAS.331..649B}, hereafter BE) conducted analysis of the astrometric microlensing events in the context of Gaia mission. 
Based on their simulation of microlensing events as seen in Gaia, they concluded that only the most nearby lenses (closer than 1 kpc) will get high quality astrometric data and will have their masses computed from Gaia data alone. However, since 2002 the design and specifications of the Gaia mission have changed significantly. 
In particular, the number of observations per object have decreased by a factor of two, therefore the real Gaia's performance in BE experiment would be even worse. In our work we have simulated the real Gaia astrometric data and combined it with the high quality photometry from the ground, a solution postulated already in BE to improve the scientific yield of the Gaia mission in the field of microlensing. Moreover, we have considered only massive lenses, in particular, a black hole candidate PAR-02, which naturally have larger amplitudes of astrometric deviation than typical lenses. Moreover, we derive microlensing parallax primarily from photometric data, therefore we can extend our sensitivity to lenses at all distances in the Milky Way. 

Our example event, PAR-02, had a baseline magnitude of about 15.45 in OGLE $I$-band. This is a typical brightness for sources in the Galactic bulge coming from the population of Red Clump giants. The distances to such sources can be safely assumed as around 8 kpc and knowing the distance parameter allows us to compute the relative parallax $\pi_{rel}$, hence the distance to the lens. 
It remains correct that a significant fraction of the sources among OGLE events in the bulge are from the Red Clump region (\eg \citealt{2015ApJS..216...12W}).
Moreover, for nearby lenses and distant sources the relative parallax $\pi_{rel}$ is the largest, hence the microlensing parallax is more easily detected and robustly measured.

In case of events in the Galactic disk we note that some parts of the Milky Way will be observed by Gaia with the cadence higher than average, reaching in some parts even 200 observations over 5 years of the mission. Any microlensing events in these regions (at $\pm$45 deg ecliptic latitude, Fig. \ref{fig:scanning}) will have better chances for more accurate astrometric parameter determination due to a better sampling. On the other hand, the average number of points for bulge will be around 50. Therefore, our examples, for which the simulations were performed for the bulge region with 92 Gaia observations, can be placed somewhere in the middle of the overall sampling for disk and bulge, where the probability of the microlensing detection is the highest (\citealt{2008ApJ...681..806H}, \citealt{2010GReGr..42.2047M}). 

The varying observed magnitude of the event affects the photometric and astrometric data accuracy as seen in Table \ref{tab:astrom_err}.
Gaia astrometric precision {\it per epoch} degrades fairly quickly with the observed magnitude and this is the main factor of degradation in the mass lens determination in the events with OGLE photometry and Gaia astrometry (Fig. \ref{fig:simulmag-mass}).
The photometric precision varies with magnitude as well, but the scatter still remains good at a level of few per cent for the faintest cases considered here. Moreover, with relatively high cadence (at least once per couple of nights), the photometric light curve can still provide high signal-to-noise data for microlensing events even at faint magnitudes. 


Despite the Gaia astrometric motion curves will only be available with the final Gaia data release (around 2022), the microlensing events on-going during the mission should be intensively observed in order to collect enough information to combine with the Gaia data. The most crucial is the microlensing parallax parameter. The OGLE bulge events will be covered well by OGLE itself, however, for the shorter events it might be necessary to use Spitzer for simultaneous observations and microlensing parallax determination. All disk events detected by Gaia itself (or other Galactic Plane surveys) should be observed frequently to cover their light curve evolution densely enough to detect the parallax signal. Moreover, the spectroscopic determination of the source distance is necessary (e.g. \citealt{2016A&A...595L..11S}), especially for the sources that do not reside in the bulge. 

In order to increase the chances of detecting a lensing black hole, many more events should have their masses derived, therefore we should reach to fainter events, which are significantly more common. The proposed, but yet unsuccessful, new astrometric mission, Theia \citep{2016SPIE.9904E..2FM}, daughter of Gaia, would be able to provide 4 per cent mass measurements for PAR-02-like events (see dash-dotted line on Figure \ref{fig:masserror}). Such data is required to reach to the faintest events (which dominate in numbers) and also lower masses of black holes.

\section{Conclusions}
In this work we tested the potential of Gaia in detecting stellar mass black holes among the lenses in microlensing events when astrometric motion curves from Gaia are combined with good quality photometry from the ground. 
We analysed one archival event, OGLE3-ULENS-PAR-02/OGLE-2006-BLG-095 (PAR-02), as it is a good candidate for a stellar black hole\citep{2016MNRAS.458.3012W}. The photometric data alone provided part of the information about the lens (including the microlensing parallax), though not sufficient for the mass determination. 
Basing on the realistic expectations for Gaia astrometric performance, we simulated motion curve data for PAR-02-like events as if they were observed by Gaia over its 5 year mission. Simultaneous fitting of the photometry and astrometry resulted in the probability distributions for the event parameters, combination of which led to lens mass and distance determination. We found that if the mass of the lens in PAR-02 was $\sim$$10~M_{\odot}$, it would be measured by Gaia with accuracy of about 25 per cent, however, at the level of brightness of PAR-02 (baseline $I$=15.5 mag or $G$=16.8 mag), the degeneracy between two solutions would not be broken with addition of the astrometry. 
For similar, but brighter events, the mass measurement would be more accurate with 12 and 8 per cent for baselines of $I$=14 mag and $I$=13 mag, respectively (equivalent of about $G$=15.3 and $G$=14.3 mag, respectively) and in case of $I$=13 mag, the input solution was found uniquely without the degeneracy. 

We conclude that despite the limitation to the brightest events only, combination of unprecedented accuracy of the Gaia astrometry and high cadence, very precise and dense photometric measurements from OGLE or ground-based follow-up, will lead to the detection of a few stellar origin, isolated black holes in the Galaxy.
The sensitivity increases with increasing mass of the lens and GW-like black holes (M$>$20 \msun), if there is enough of them in the Galaxy, should be detected and have their masses measured with $<$15 per cent accuracy.

\section*{Acknowledgements}
We would like to thank Drs Vasily Belokurov, Andrzej Udalski, Jan Skowron, Przemek Mr{\'o}z for their help at various stages of this work and valuable discussions. We would also like to thank Dr Simon Hodgkin for the careful reading of the manuscript and many helpful comments. We acknowledge the entire Gaia DPAC, who is running the Gaia mission, in particular for preparation of the Gaia Observations Forecast Tool (https://gaia.esac.esa.int/gost/). The authors thank the anonymous referee for their comments which significantly improved the manuscript.

KR acknowledges support from the Polish National Centre grant OPUS no. 2015/17/B/ST9/03167 to {\L}W.
This work was partly supported by the Polish National Centre grant HARMONIA no.2015/18/M/ST9/00544 to {\L}W.



\bibliographystyle{mnras}
\bibliography{bibliography} 

\bsp	
\label{lastpage}
\end{document}